# Crystal structure, lattice dynamics and superexchange in $MAgF_3$ 1D antiferromagnets (M = K, Rb, Cs) and $Rb_3Ag_2F_7$ Ruddlesden-Popper phase

Kacper Koteras,[1]* Jakub Gawraczyński,[1] Gašper Tavčar,[2] Zoran Mazej,[2] Wojciech Grochala[1]

[1] Centre of New Technologies, University of Warsaw, S. Banacha 2C, 02-097 Warsaw, Poland

[2] Jožef Stefan Institute, Jamova cesta 39, 1000 Ljubljana, Slovenia

*Dedicated to Prof. M. Niezgodka at his 70th birthday*

*e-mail: k.koteras@cent.uw.edu.pl

## Abstract

With the use of lattice dynamics calculation within hybrid HSE06 framework we were able to understand vibrational spectra of $MAgF_3$ (M = K, Rb, Cs) compounds. Comparative theoretical study uncovered lack of monotonicity in calculated optical phonons associated with Ag-F stretching modes which can be explained through an interplay of Lewis acidity of M(I) cation and its size. We confirm the tetragonal unit cells of $MAgF_3$ (M=Rb, Cs) at room temperature. We also theoretically predict an orthorhombic $RbAgF_3$ polymorph as a ground state at low temperature. However, we were not able to detect it by the means of low temperature PXRD (at 80 K) nor low temperature Raman (at 154 K) due to a number of constraints. We also describe a novel Ruddlesden-Popper phase of $Rb_3Ag_2F_7$ which can be regarded as quasi-0D system, where superexchange coupling constant between the nearest Ag(II) centres reaches an impressive value of –240.2 meV.

## 1. Introduction

Systems composed of silver(II) and fluorine are in many ways exceptional.[1] Silver(II) is so reactive that fluorine is the only element, which may form thermodynamically stable compounds with it. The binary $AgF_2$, on its own right, hosts many exciting phenomena such as strong mixing of Ag(d) and F(p) states,[2] appreciable magnetic superexchange in 2D,[1] remarkable covalence of Ag–F bonds[3] and electronic structure at the verge of valence instability.[3] These features render $AgF_2$ similar to copper(II) oxides and $BaBiO_3$, which are precursors of two important families of superconductors. Said similarity places Ag(II) fluorides in the focus of solid state chemistry, physics and materials science. Introducing the third element to the chemical composition further broadens the playground and provides us with a multitude of interesting physical phenomena to be studied.[4–6] The simplest modifications of $AgF_2$ are exemplified by ternary fluoroargentates(II), $M(I)AgF_3$ and $M(I)_2AgF_4$. While $M_2AgF_4$ systems have already been thoroughly studied,[7–9] the $MAgF_3$ ones (particularly M = Rb and Cs) remain explored unsatisfactorily.

Alkaline ternary fluoroargentates(II) of a general formula $MAgF_3$ (M = K, Rb, Cs) adopt a distorted perovskite structure, with alkaline atom substantially impacting level of distortion (Figure 1). Generally, smaller the alkali metal cation, the more distant from ideal cubic lattice is the resulting structure. The early study of these compounds was conducted half a century ago, when quality of laboratory X-ray sources was much worse than nowadays. An orthorhombic (*Pnma*) structure was then assigned to $KAgF_3$ and tetragonal (*I*4/*mcm*) ones to $RbAgF_3$ and $CsAgF_3$ (based on Guinier pattern analysis).[10] Although data quality and purity of the samples were far from good, these are the only structures of $RbAgF_3$ and $CsAgF_3$ reported so far. Only in 2013, it was possible to substantially improve purity of $KAgF_3$ specimen, which permitted more sophisticated analyses to be conducted.[11] However, syntheses of $RbAgF_3$ and $CsAgF_3$ have not yet been improved. There is even less research regarding other alkali fluoroargentates(II). $NaAgF_3$ is proposed to be a monoclinic or triclinic $NaCuF_3$-related post perovskite[12] while $LiAgF_3$ has not been prepared despite many attempts.[13]

The more recent study of $KAgF_3$ has showed that this compound exhibits a structural phase transition at 230 K which is of order/disorder type. The low-temperature phase is in *Pnma* symmetry, in agreement with the early study, but the high-temperature phase adopts *Pcma* symmetry and twice smaller unit cell.[11] Moreover, experimental evidence of complex magnetic ordering below 66 K was also obtained.[6] Magnetic properties of $RbAgF_3$ and $CsAgF_3$ were only scarcely described in the structure determination paper.[10] The main obstacle while analysing magnetic properties of $RbAgF_3$ and $CsAgF_3$ was that that samples were contaminated with $M_2AgF_4$ and $M_3AgF_7$ (M = Rb, Cs) phases. The $M_2AgF_4$ phases are ferromagnetic,[4] thus obscuring magnetic signals from antiferromagnetic $MAgF_3$ ones.

A recent theoretical study of $MAgF_3$ compounds assumed the previously proposed crystal structures and it mostly focused on their magnetic properties. It seems that all compounds from the $MAgF_3$ family can be regarded as quasi-1D antiferromagnets, with strong AFM intra-chain superexchange, and much weaker FM inter-chain one.[5] The calculated intra-chain superexchange constant $J_{1D}$ is impressively large, varying between –113 meV for $KAgF_3$, via –144 meV for $RbAgF_3$, and reaching –161 meV for $CsAgF_3$. In all cases the inter-chain superexchange constants do not exceed +10 meV. Compounds exhibiting such strong magnetic interactions are rare and are of interest for solid state physics; when appropriately modified, they could be used in spintronics e.g. for data storage.

The above-mentioned theoretical study[5] puts in question the tetragonal unit cell of $RbAgF_3$; authors have noticed that an orthorhombic polymorph of $RbAgF_3$ has lower total energy than the tetragonal one. One possible reason for this is that experimental study[10] has been conducted at 293 K, while theoretical one corresponds formally to 0 K. Another reason might be the poor quality of old X-ray data, which did not permit to detect small orthorhombic distortion of the tetragonal cell.

In our recent work, we have studied lattice dynamics of $KAgF_3$ in order to understand its phase transition and to assign vibrational modes.[14] Very good agreement between theory and experiment was reached. Phonon calculations also proved stability of the low temperature orthorhombic polymorph of $KAgF_3$. As up to now, no lattice dynamics studies of $RbAgF_3$ and $CsAgF_3$ were performed, we have decided to conduct a joint experimental/theoretical study of these compounds. The goal of this study is to expand our recent work on $KAgF_3$ by assigning vibrational spectra of $RbAgF_3$ and $CsAgF_3$ and to gain better understanding of $RbAgF_3$ crystal structure. Moreover, by comparing the data for all three members of the $MAgF_3$ family, we aimed to detect trends in chemical bonding and vibrational structure across the series.

As a bonus, we have been able to determine crystal structure of $Rb_3Ag_2F_7$ Ruddlesden-Popper phase, support the space group assignment using theoretical density functional theory (DFT) calculations, and calculate the magnetic superexchange constant for discrete $Ag_2F_7^{3-}$ dimers present in its structure..

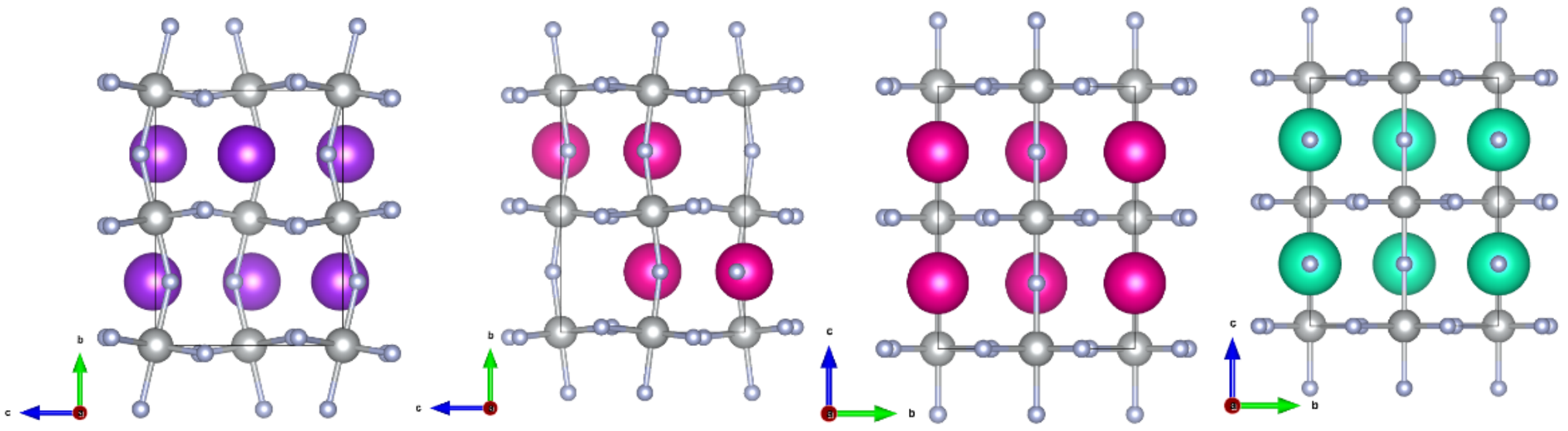

Figure 1. Crystal structures of (from left): $KAgF_3$ (in low temperature *Pnma* space group), $RbAgF_3$ (in theoretically predicted *Pnma* space group), $RbAgF_3$ (in experimental *I*4/*mcm* space group), and $CsAgF_3$ (*I*4/*mcm*). Colour coding: grey – silver, light blue – fluorine, violet – potassium, pink – rubidium, cyan – caesium.

## 2. Experimental

$KAgF_3$ syntheses were conducted in three different ways: (1) annealing of ground mixture KF and $AgF_2$ at 300°C for four days in a nickel reactor under $F_2$ atmosphere, (2) thermal decomposition of $KAgF_4$ maintaining dynamic vacuum at 420°C, (3) fluorination of $KAg(CN)_2$ with stoichiometric quantity of fluorine gas at 300°C. $RbAgF_3$ sample was synthesized by annealing of hand ground RbF and $AgF_2$ mixture. This was conducted in a nickel boat in a nickel autoclave for 24 h at 310°C. After that time resultant mixture was ground again and annealed in the same reaction for additional 24 h. $CsAgF_3$ synthesis was conducted by annealing of hand ground mixture of CsF and $AgF_2$ (same conditions as $RbAgF_3$) or thermal decomposition of $CsAgF_4$ (same conditions as $KAgF_3$). Better purity $CsAgF_3$ sample was obtained from the first approach.

Infrared measurements were carried out on Bruker Vertex 80 V vacuum spectrometer using powdered samples placed on cleaned and dried HDPE windows.

Raman spectra for $KAgF_3$ and $RbAgF_3$ were obtained on a Horiba Yvon LabRam-HR Raman micro-spectrometer with 632.8 nm He-Ne exciting beam. Due to technical issues with abovementioned spectrometer, spectrum for $CsAgF_3$ was obtained on Renishaw inVia micro-spectrometer with 785 nm excitation beam. Because of contamination of $RbAgF_3$ and $CsAgF_3$ samples with small amounts of $Rb_2AgF_4$ and $Cs_2AgF_4$ double perovskites, and triple perovskite $Rb_3Ag_2F_7$ Raman spectra varied somewhat from one measurement spot to another. Results chosen for further analysis were crosschecked with published data to confirm that we indeed measured spectra for $MAgF_3$ rather than $M_2AgF_4$ crystallites.[14–16]

Low temperature Raman spectra were measured in a custom setup described in our recent work[14] using Horiba Yvon LabRam-HR Raman micro-spectrometer with 632.8 nm He-Ne exciting beam. Setup consists of FEP tube equipped with capillary holder, dry air cooled by $LN_2$ and a thermocouple.

Room temperature powder X-ray diffraction measurements were carried out on Bruker D8 Discover diffractometer, using CuKα radiation. Sample was placed in heat sealed in prefluorinated 0.3 mm quartz capillary. Measurement 2σ range spanned from 6° to 90°, with step size of 0.014°. Acquisition time used here was equal to 16 h for whole measurement in four consecutive runs to determine if sample reacted with the capillary. Low temperature measurements were conducted on Oxford Diffraction Xcalibur 2 single crystal diffractometer also using CuKα radiation. Resolution of these measurements is worse than room temperature measurements, because detectors in these diffractometers are usually placed much closes to the sample. Using this setup had an advantage of reaching 80 K with high accuracy.

Room temperature powder data was refined using Rietveld method[17] in Fullprof software suite.[18] Full set of profile (including asymmetry) and atomic (positions and isotropic thermal factors) parameters were refined. Absorption correction factor of μR = 1.1 was used.

## 3. Computational details

Lattice Lattice dynamics of $MAgF_3$ (M = K, Rb, Cs) compounds was modelled using hybrid functional HSE06,[19] based on Perdew-Burke-Ernzerhof DFT-GGA functional revised for solids (PBEsol)[20] with 25% Hartree Fock exchange energy. For all calculations, projector-augumented-wave method[21,22] was used, as implemented in VASP 5.4.4 code.[23–25] Standard VASP pseudopotentials were used for description of core electrons.

All structures were optimized by performing a full relaxation of unit cell parameters and atomic coordinates. Antiferromagnetic (AFM) ordering was assumed for all structures in agreement with previous results (i.e. AFM ordering within chains propagating along the crystallographic ***c*** direction, and ferromagnetic (FM) one within sheets parallel to ***ab*** planes).[5,26] The cut-off energy of the plane wave basis set was equal to 520 eV with a self-consistent-field convergence criterion of $1 \cdot 10^{-7}$ eV. The k-point mesh spacing used in calculations was equal to 0.25 Å$^{-1}$. All structures were optimized until the forces acting upon each atom were smaller than 0.001 eV/Å.

Phonon calculations were conducted using finite differences method implemented in VASP. Results were then processed using Phonopy package in order to extract irreducible representations and dispersion curves, as VASP provides frequencies only in the centre of the first Brillouin zone (Γ point).[27] Animated pictures of normal modes, also generated by Phonopy, were used to assign mode types. Calculations were conducted for single unit cells, in order to limit resource usage. Phonon band structures were calculated for standardized k-path.[28] For tetragonal *I*-centered cells, primitive cells (Z=2) were used for band structure and irreproducible representations calculations.

Magnetic superexchange constants were calculated using broken symmetry method mapping calculated energy values on spin Hamiltonian in a form:

$$H = -0.5 \sum_{i,j} J_{ij} s_i s_j$$

Total energy values for each magnetic configuration were obtained by the means of GGA+U by conducting single point calculations using geometry of ground state configuration. The k-point mesh spacing used in calculations was equal to 0.16 Å$^{-1}$. The cut-off energy of the plane wave basis set was equal to 800 eV with a self-consistent-field convergence criterion of $1 \cdot 10^{-7}$ eV.

# 4. Results and discussion

## 4.1. Theoretical phonon spectra.

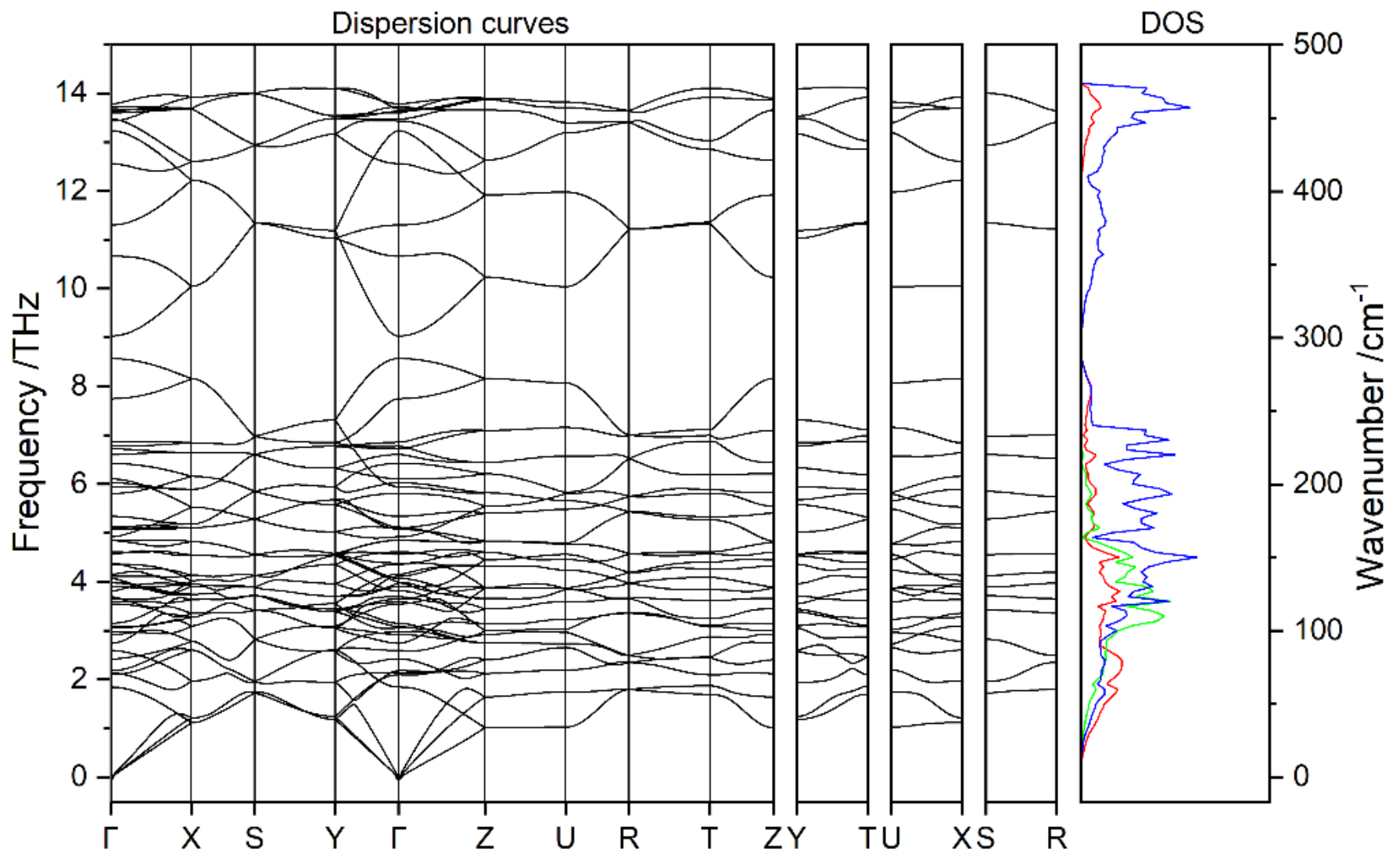

Figure 2. Calculated phonon dispersion curves and (atomic) phonon density of states for $KAgF_3$ (blue – F, red – Ag, green – K)

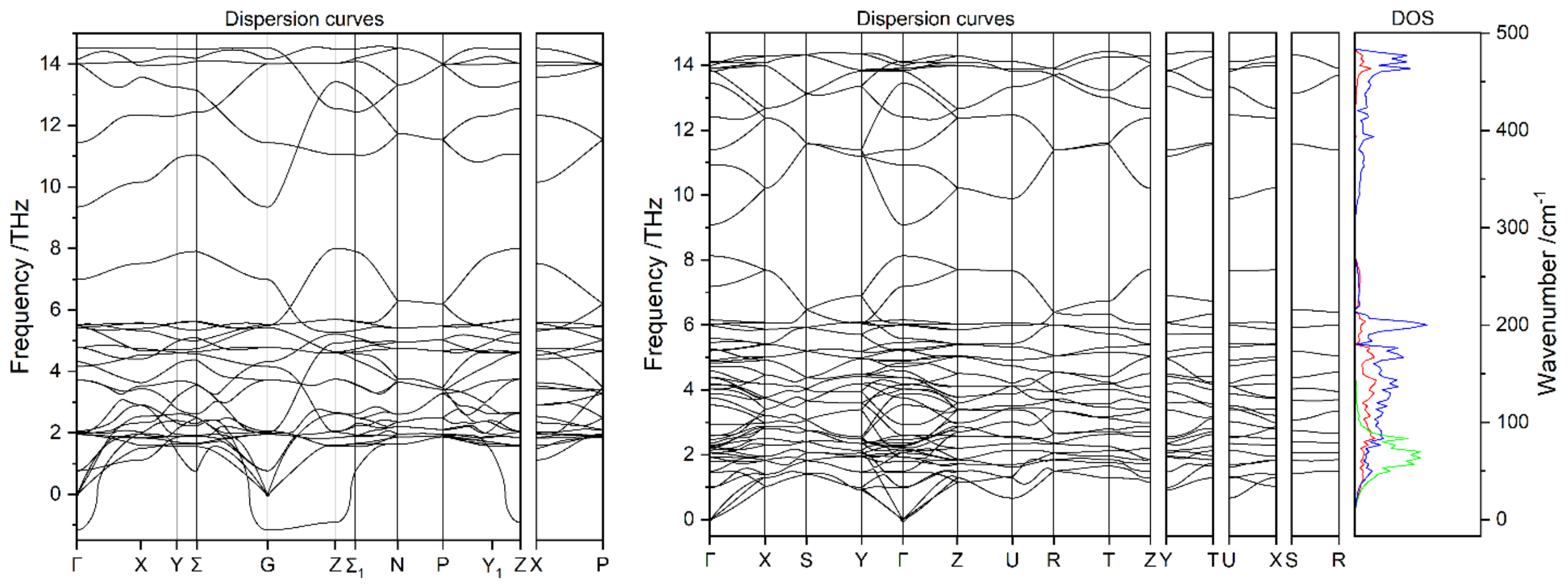

Figure 3. Calculated phonon dispersion curves for tetragonal $RbAgF_3$ (left), and combined graph of dispersion curves and phonon density of states for orthorhombic $RbAgF_3$ (right) (blue – F, red – Ag, green – Rb)

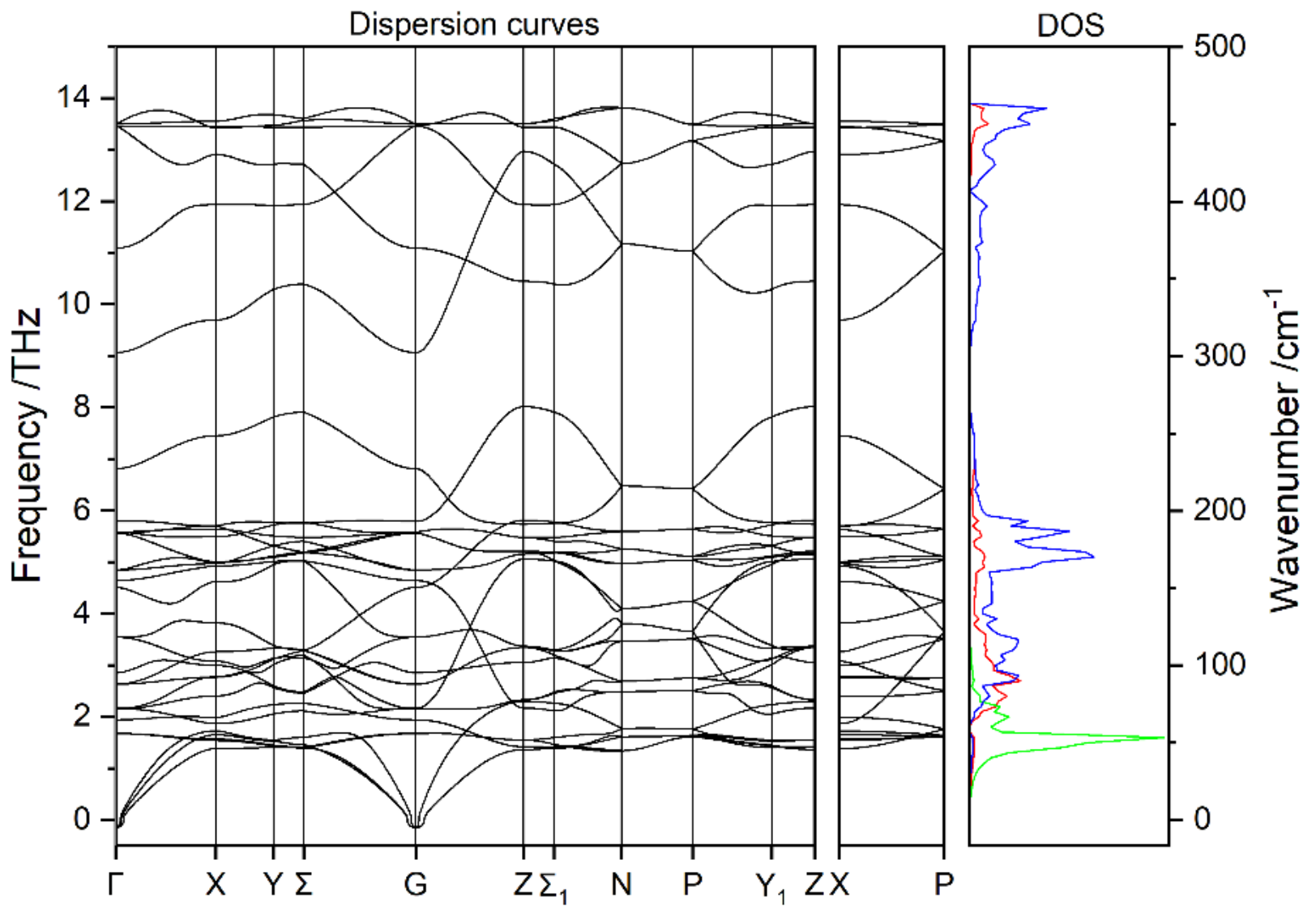

Figure 4. Calculated phonon dispersion curves and (atomic) phonon density of states for tetragonal $CsAgF_3$ (blue – F, red – Ag, green – Cs)

Dispersion curves calculated for $KAgF_3$ using HSE06 dataset and Phonopy suite (Figure 2), remain in a good agreement with those computed recently utilizing a commercial PHONON software.[14] All features of phonon density of states were also successfully reproduced. Therefore, for the remaining systems their phonon spectra were computed using Phonopy.

Our lattice dynamics calculations using published, tetragonal structure of $RbAgF_3$ show significant dynamic instability at the centre of the first Brillouin zone (Γ) but also in other special points (Figure 3, top). The presence of an imaginary branch alone is the proof that the tetragonal structure cannot be the ground state of this system. Due to that, we have used a modified model in which we have allowed for the presence of kinked [$AgF_3^-$] chains similar to those typical of $KAgF_3$. Full geometry relaxation using HSE06 functional yielded a dynamically stable structure (Figure 3, bottom), that had 11.7 meV per unit cell (Z=4) lower total energy than the starting tetragonal one. That energy difference corresponds roughly to a temperature of 34 K, so it may be deduced than the orthorhombic form corresponds to a low temperature polymorph. As a result of larger ionic radius, the [$AgF_3^-$] chains in theoretically calculated $RbAgF_3$ are less kinked than in calculated $KAgF_3$ (165.5° vs. 153.3°). This effect can also be seen by analysing density of states graph. Vibrational states of Rb are more localized than those of K signalling a larger freedom of potassium ions in $KAgF_3$ for rattling.

Phonon dispersion curves calculated for a primitive cell of tetragonal $CsAgF_3$ (Figure 4) confirm dynamic stability of this structure. A small departure of one mode from null value (*img* 5 $cm^{-1}$) stems from imperfect diagonalization of Hessian matrix, and may be considered artefactual. Localization of alkali metal vibrational states is in $CsAgF_3$ even more apparent meaning that caesium, with its large ionic radius, sits tightly in interstitial positions. Consequently, Ag-F-Ag angles present here are straight (*i.e.* 180°). This comes from the fact that among alkali metal cations, the Cs(I) cation best fulfils the geometric condition of structural stability of a perovskite structure.

Due to similarity in $MAgF_3$ (M = K, Rb, Cs) structures, normal modes involving similar atomic displacements can be found in each spectrum (Figure 5). In the highest wavenumber region (425-475 $cm^{-1}$) modes calculated for $RbAgF_3$ corresponding to stretching vibrations of the Ag–F bond network exhibit higher energies than those for $KAgF_3$ and $CsAgF_3$. This is an interesting feature which may be explained by considering that there are two main factors in play in the M = K, Rb, Cs series. One factor is related to the Lewis acidity of M(I) cations. K(I) cation having the largest Lewis acidity in the series leads to its strongest interaction with $F^-$ ions and thus to some weakening of the Ag-F bonds. Volume of the M(I) cation is the other factor, ‡[29] which – particularly for the largest Cs(I) cation, may also induce weakening of the Ag–F bonds due to "negative pressure" exerted by this cation. It seems that the problem of two mutually opposing factors has an optimum and the Rb(I) salt shows the highest frequencies of Ag–F stretching modes.

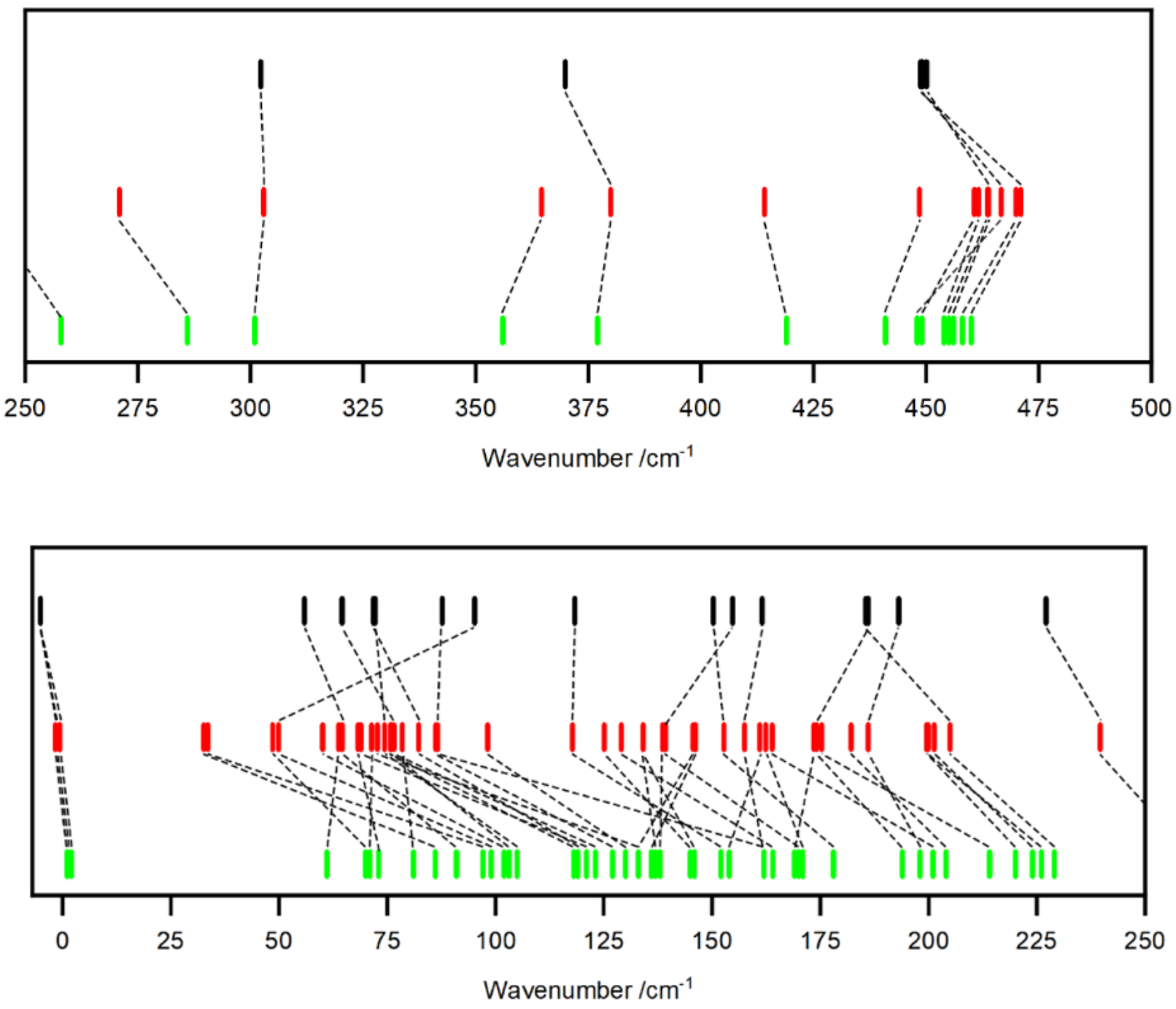

Figure 5. Calculated wavenumber of normal modes at Γ for orthorhombic $KAgF_3$ (green), orthorhombic $RbAgF_3$ (red) and tetragonal $CsAgF_3$ (black). Dashed lines connect corresponding modes with similar forms of vibrations. Note that the number of modes computed for $CsAgF_3$ in its primitive cell is twice smaller than those for orthorhombic K and Rb ones, hence some modes do not have counterparts in the spectrum of Cs derivative.

Midrange [$AgF_6$] octahedron stretching modes (300-425 $cm^{-1}$) are generally close in energy for all three compounds. However, finding the modes corresponding to one another in the spectral region below 250 $cm^{-1}$ (*e.g.* lattice modes) is not that straightforward, because of significant admixture of heavy alkali metal vibrations in each mode. Generally normal modes found for $RbAgF_3$ are softer than analogous modes of $KAgF_3$ in agreement with the *ca.* twice reduced oscillator mass for the latter. Corresponding modes of $CsAgF_3$ have comparable values to those for Rb analogue.

### 4.2. Assignment of vibrational spectra.

Our analysis of powder x-ray pattern shows that at room temperature $RbAgF_3$ adopts tetragonal structure. As stated above, this is supposedly a high-temperature form, with orthorhombic one corresponding to the genuine ground state at T → 0 K. Since our spectroscopic measurements were conducted at room temperature, we have used phonons calculated for tetragonal polymorph for an interpretation of IR and Raman data.

Infrared spectrum measured for $RbAgF_3$ consists of 8 bands, while Raman measurement yielded 6 bands (Figure 7). Group theory analysis for that compound in a primitive cell of the *I*4/*mcm* space group yields 21 distinct normal vibrational modes, including some degenerated ones ($\Gamma_{acoustic} = A_{2u} + E_u$; $\Gamma_{optic} = A_{1g} + 2A_{2g} + 3A_{2u} + B_{1g} + 2B_{1u} + 2B_{2g} + 5E_u + 3E_g$); here, there are 7 Raman active ($A_{1g}$, $B_{1g}$, $2B_{2g}$, $3E_g$) and 8 IR-active modes ($3A_{2u}$, $5E_u$). The remaining modes are silent. Measured spectra for $CsAgF_3$ in *I*4/*mcm* symmetry consist of 9 infrared bands and 4 Raman bands. Obviously, the group theory predicts identical result as for the Rb salt.

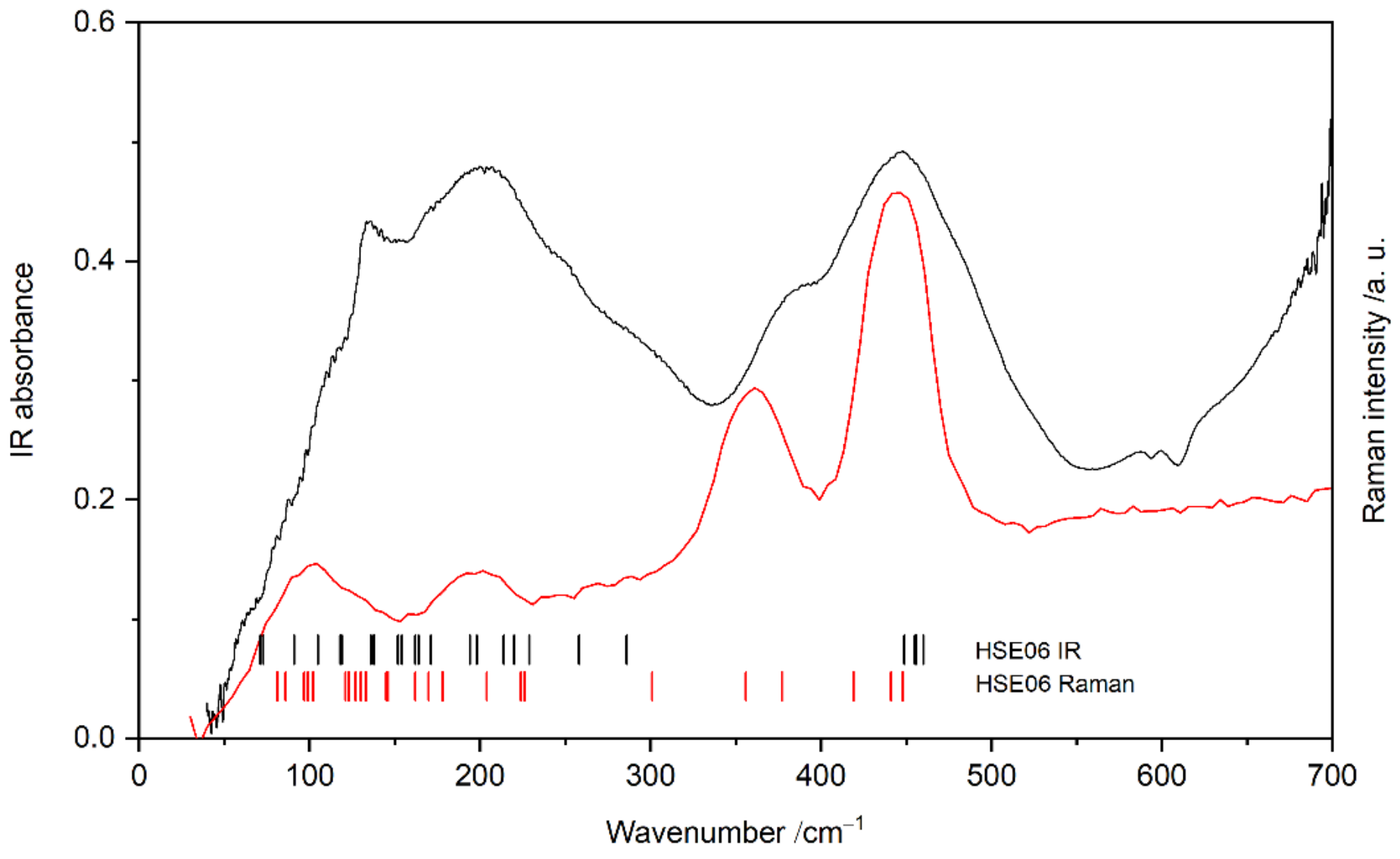

Figure 6. FIR (black) and Raman (red) spectra of $KAgF_3$ at room temperature together with marks indicating the computed wavenumbers.

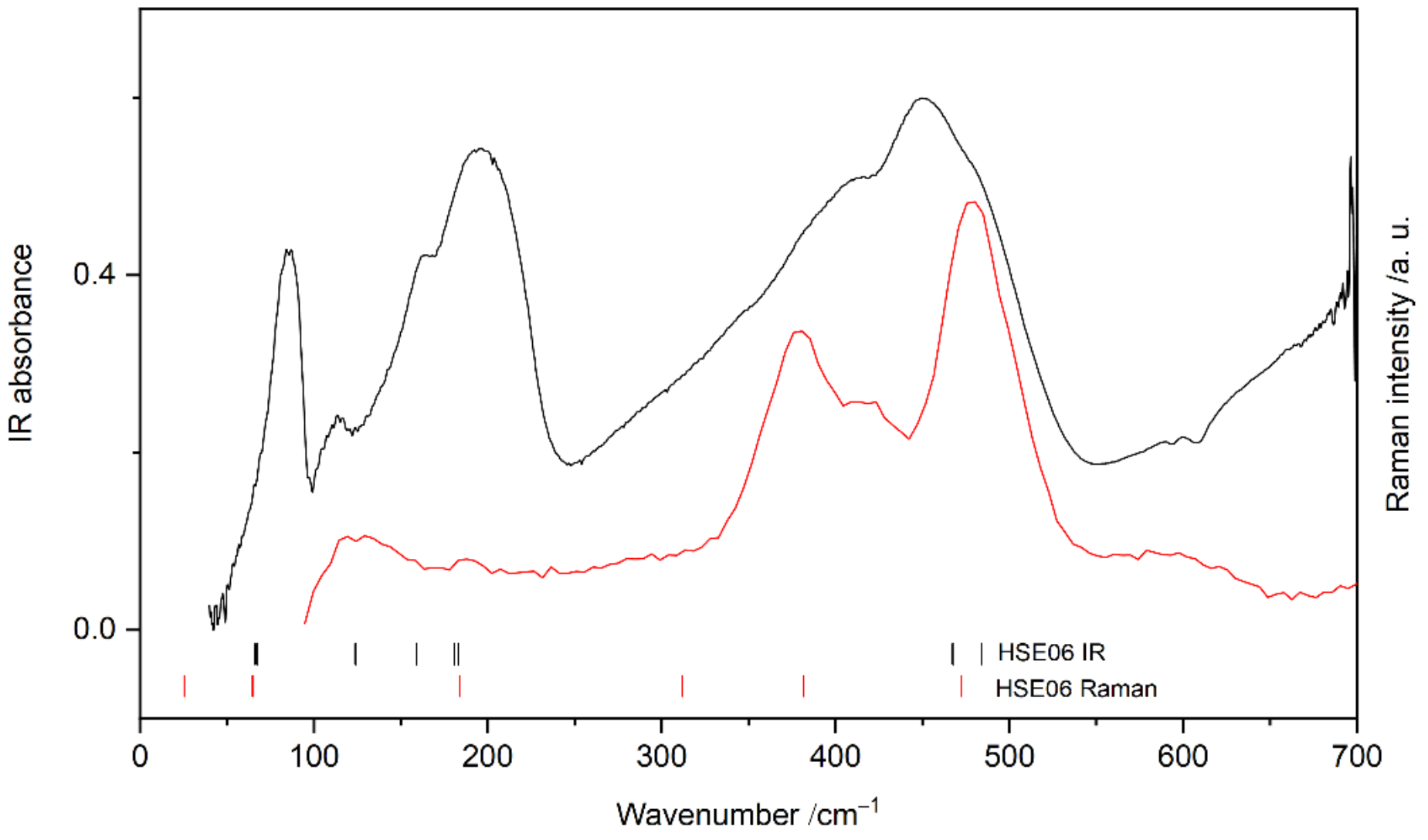

Figure 7. FIR (black) and Raman (red) spectra of $RbAgF_3$ at room temperature together with marks indicating the computed wavenumbers.

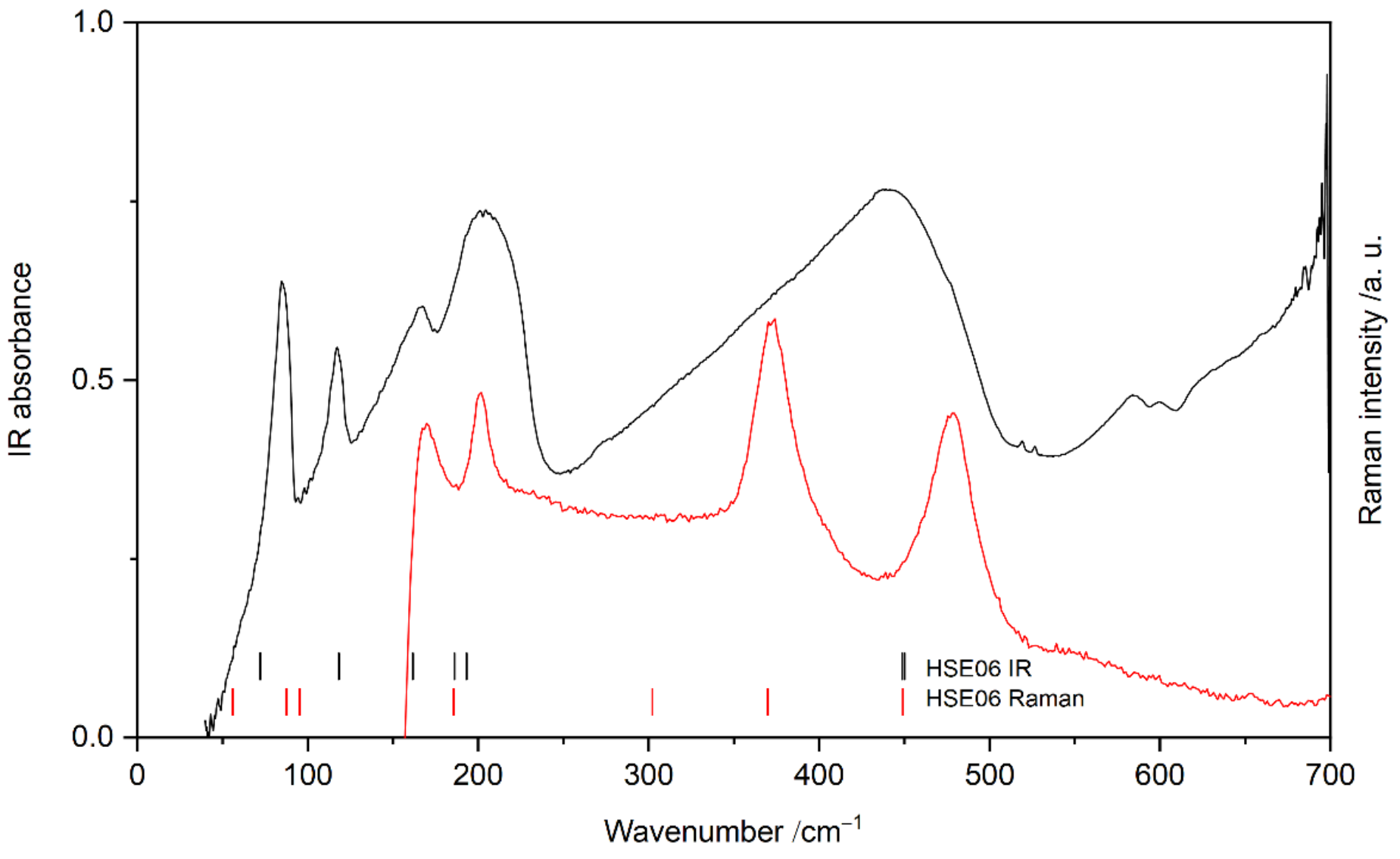

Figure 8. FIR (black) and Raman (red) spectra of $CsAgF_3$ at room temperature together with marks indicating the computed wavenumbers. The Raman spectrum for Cs salt shows an abrupt decline in intensity below 170 $cm^{-1}$ due to a notch filter.

$KAgF_3$ spectra were described in detail in our recent work based on DFT+U and HSE06 calculations.[14] Here, in Figure 6, we show the experimental spectra overlapped with theoretically predicted modes using highly accurate hybrid HSE06 functional (without any scaling factor), so that they may be compared with those measured for its heavier analogues.

Out of eight infrared bands present in $RbAgF_3$ spectrum (Figure 7), six were successfully assigned to fundamentals and two to combination modes (using group theory constraints for selection rules) or impurities (Table 1). The highest wavenumber band in infrared spectrum is placed at 460 $cm^{-1}$ and was assigned to $E_u$ mode representing Ag–F stretching within [$AgF_2$] planes. Three midrange bands at 200, 162 and 113 $cm^{-1}$ correspond to bending modes assigned to $A_{2u}$ [$AgF_2$] plane buckling, $E_u$ [$AgF_2$] in plane bending and $E_u$ [$AgF_6$] octahedron bending respectively. The lowest wavenumber bands (85 and 72 $cm^{-1}$) are rubidium lattice modes. The Raman spectrum features nine six of which three were assigned to fundamentals. The band at 477 $cm^{-1}$ has the largest intensity; it is assigned to [$AgF_6$] octahedron stretching mode. The second highest intensity band located at 378 $cm^{-1}$ was assigned to [$AgF_2$] plane stretching. In the lower wavenumber range, there is one band at 189 $cm^{-1}$ which was assigned to [$AgF_6$] octahedron bending mode.

Infrared spectrum of $CsAgF_3$ (Figure 8) consists of nine bands, six of which were successfully assigned to corresponding fundamental modes and three either to combination/overtone modes or impurities (Table 2). The highest wavenumber infrared fundamental at 448 $cm^{-1}$ falls very close to two predicted normal modes, leaving some ambiguity in assignment; it can either be a [$AgF_2$] plane stretching or [$AgF^+$] chain stretching mode. Similarity to the Rb and K salt suggest preference for the former assignment. Band seen at 206 $cm^{-1}$ is [$AgF_2$] plane buckling mode, 163 $cm^{-1}$ is [$AgF_6$] octahedron bending mode, and 117 $cm^{-1}$ is [$AgF^+$] chain bending mode. Two lowest energy bands at 86 and 80 $cm^{-1}$ are Cs lattice modes. Raman spectrum was deconvoluted to four major bands, of which three were assigned to fundamentals. Stretching modes are at 477 $cm^{-1}$ ([$AgF_6$] octahedron stretching) and 372 $cm^{-1}$ ([$AgF_2$] plane stretching). The third band at 202 $cm^{-1}$ originates from [$AgF_6$] octahedron bending.

The correlation between theoretically predicted and measured wavenumbers for both the Rb and Cs salt is excellent, testifying that scaling factor is nearly unity (slope factor is 0.975 for Cs, 0.995 for Rb) when using the HSE06 hybrid functional (Figure 9).[14]

Table 1. Vibrational band assignment for tetragonal $RbAgF_3$ (at room temperature).

| IR /$cm^{-1}$ | Raman /$cm^{-1}$ | Type | HSE06 /$cm^{-1}$ | Symmetry | Assignment |
|---|---|---|---|---|---|
| | 602 | w | - | - | impurity |
| | 507 | sh | - | - | impurity (possible $Rb_3Ag_2F_7$) |
| | 477 | s | 472 | $B_{2g}$ | [$AgF_6$] octahedron stretching |
| 460 | | s | 467 | $E_u$ | [$AgF_2$] plane stretching |
| | 423 | w | - | - | impurity |
| 395 | | sh | - | - | combination 200+189 ($A_{2u}$ x $E_g$) or impurity |
| | 378 | s | 381 | $A_{1g}$ | [$AgF_2$] plane stretching |
| 347 | | sh | - | - | combination 162+189 ($E_u$ x $E_g$) or impurity |
| 200 | | s | 183 | $A_{2u}$ | [$AgF_2$] plane buckling |
| | 189 | m | 184 | $E_g$ | [$AgF_6$] octahedron bending |
| 162 | | sh | 159 | $E_u$ | [$AgF_2$] in plane bending |
| 113 | | w | 124 | $E_u$ | [$AgF_6$] octahedron bending |
| 85 | | s | 67 | $A_{2u}$ | Lattice (Rb out of plane) |
| 72 | | sh | 66 | $E_u$ | Lattice (Rb in plane) |

Table 2. Vibrational band assignment for tetragonal $CsAgF_3$ (at room temperature).

| IR /cm$^{-1}$ | Raman /cm$^{-1}$ | Type | HSE06 /cm$^{-1}$ | Symmetry | Assignment |
|---|---|---|---|---|---|
| 580 | | w | – | | combination 206+372 ($A_{2u}$ x $A_{1g}$) or impurity |
| | 477 | s | 449 | $B_{2g}$ | [$AgF_6$] octahedron stretching |
| 448 | | s | 450 | $E_u$ | [$AgF_2$] plane stretching |
| | | | 449 | $A_{2u}$ | [$AgF^+$] chain stretching |
| | 398 | sh | – | – | overtone (206 $A_{2u}$ x $A_{2u}$), overtone 202 ($E_g$ x $E_g$) or impurity |
| 381 | | sh | – | – | impurity |
| | 372 | s | 370 | $A_{1g}$ | [$AgF_2$] plane stretching |
| 318 | | sh | – | – | combination 117+202 ($E_u$ x $E_g$) or impurity |
| 206 | | s | 193 | $A_{2u}$ | [$AgF_2$] plane buckling |
| | 202 | m | 186 | $E_g$ | [$AgF_6$] octahedron bending |
| 163 | | m | 162 | $E_u$ | [$AgF_6$] octahedron bending |
| 117 | | m | 118 | $E_u$ | [$AgF^+$] chain bending |
| 86 | | s | 72 | $A_{2u}$ | Lattice (Cs out of plane) |
| 80 | | sh | 72 | $E_u$ | Lattice (Cs in plane) |

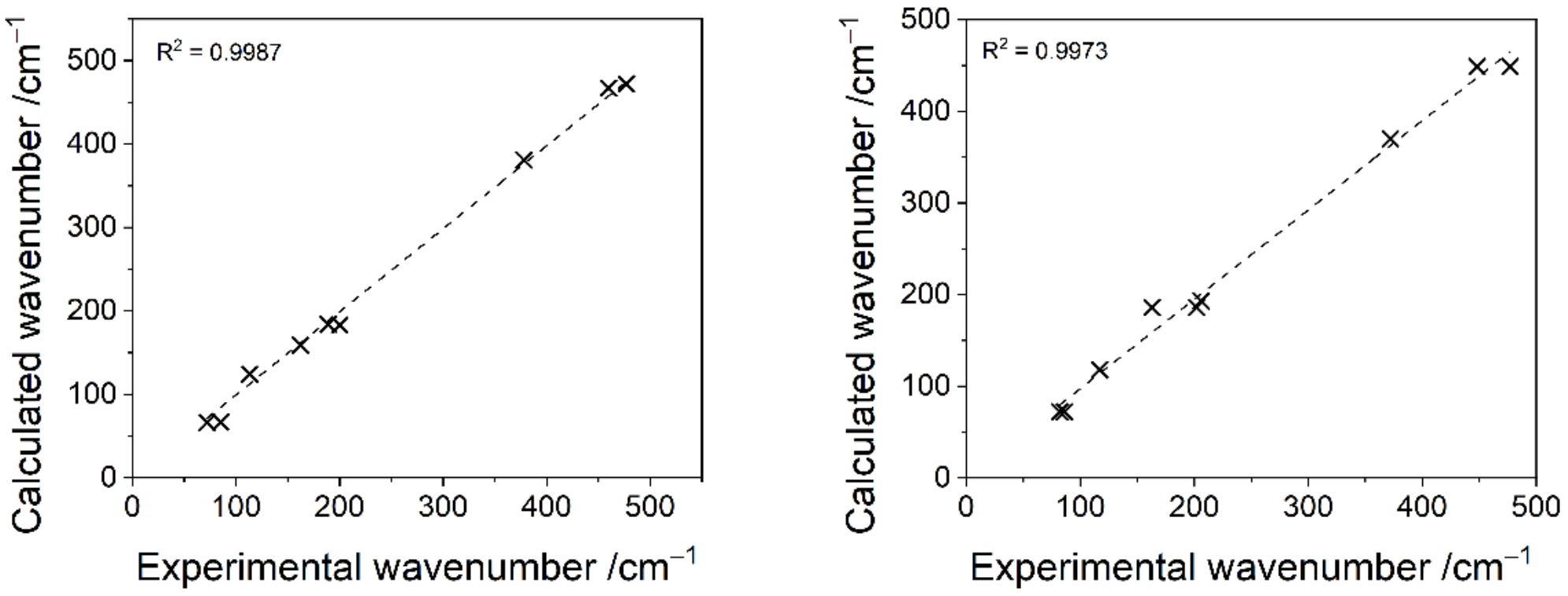

Figure 9. Correlations between theoretical and experimental wavenumbers jointly for IR and Raman spectra (left – $RbAgF_3$; right – $CsAgF_3$; black dotted line—linear regression $R^2$ depicted at the figure both bound to pass at (0,0)).

### 4.3. Vibrational spectra and x-ray diffraction pattern for $RbAgF_3$ at low temperature.

The calculated total energy at 0 K of tetragonal $RbAgF_3$ is higher than total energy of orthorhombic polymorph. In order to get further insight into a possible tetragonal/orthorhombic phase transition, low temperature Raman spectrum (at 154 K) and low temperature x-ray diffraction pattern (at 80 K) were measured.

Comparison of low temperature Raman spectrum with that measured at room temperature shows the expected stiffening of all modes (Figure 10). Two features of highest energy got shifted towards higher wavenumber range (from 378 to 382 cm$^{-1}$ and from 477 to 488 cm$^{-1}$; a similar stiffening was observed before for the K salt.[14] Spectral features are better resolved in low-T spectra, as expected. Most new bands appear in the region above 600 cm$^{-1}$, where no fundamentals can be found; they correspond to either combination modes or overtones (*cf.* ESI). These changes are not dramatic, and take place gradually while cooling down the sample (all spectra measured are available in ESI). While these measurements did not prove that a phase transition might occur, they also did not disprove such possibility, since the Raman spectra of orthorhombic $KAgF_3$ and of tetragonal $RbAgF_3$ do not differ markedly. Because of that, we have also conducted low temperature powder X-ray diffraction experiment at 80 K.

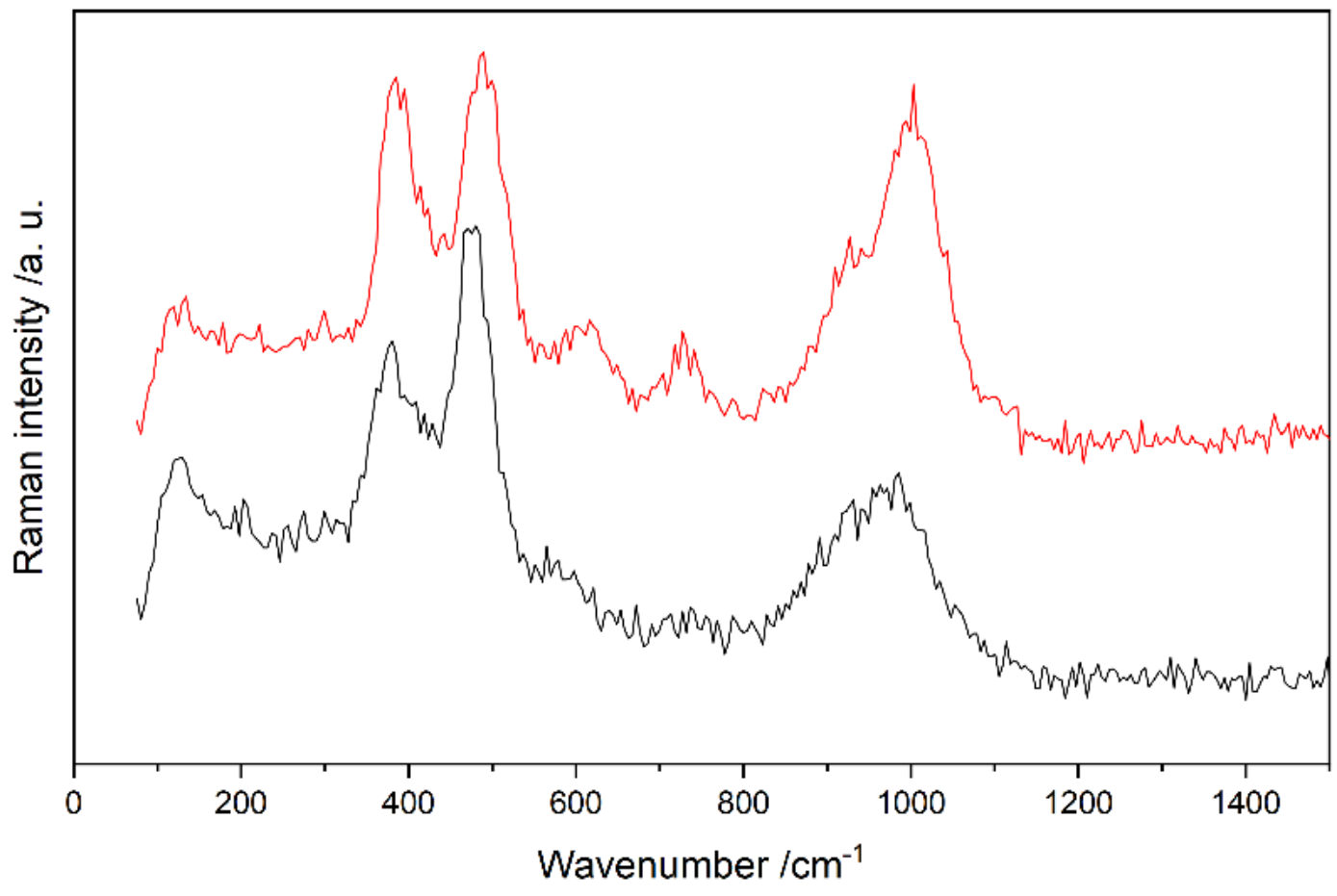

Figure 10. Raman spectrum of $RbAgF_3$ measured at room temperature (black) and at 154 K (red).

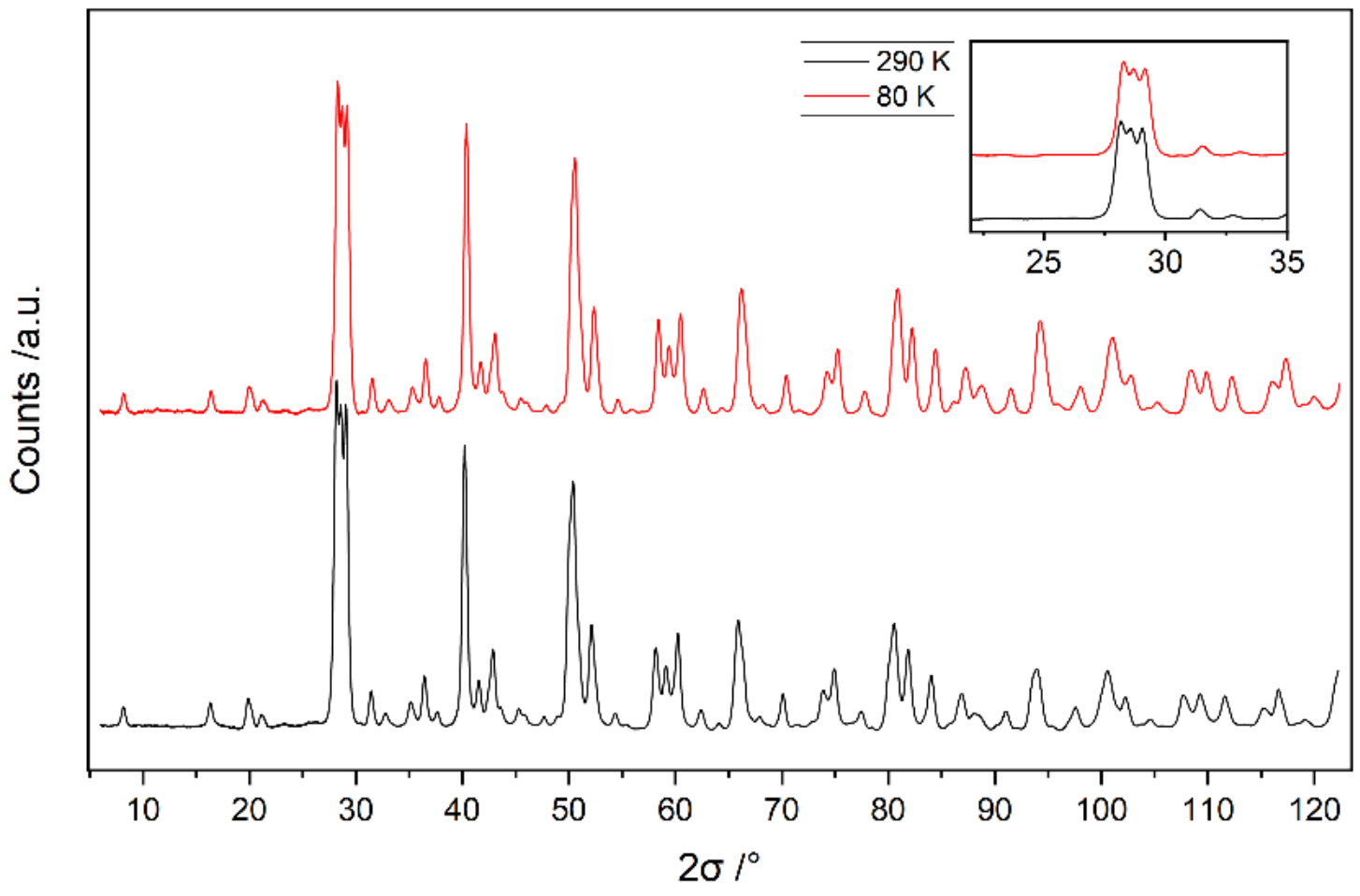

Figure 11. Background-reduced powder x-ray diffraction pattern of $RbAgF_3$ measured at room temperature (black) and at 80 K (red). Inset shows the three most intense peaks of the pattern.

We were able to reach temperature of 80 K in powder X-ray diffraction measurements (Figure 11). Resolution of that pattern is sufficient for the analysis of a possible tetragonal-to-orthorhombic transition. Theory predicts that a splitting of a tetragonal lattice parameter to two distinct values of 6.263 Å and 6.344 Å (***a*** and ***c*** vectors in the *Pnma* setting) will affect most strongly the 100 *vs.* 001, 110 *vs.* 011, 002 *vs.* 200, and related pairs of reflections. Unfortunately, they all fall in 2 theta regions where the reflections from the impurity phase are also present (see the next section). Therefore, X-ray analysis is also inconclusive with respect to a possible symmetry-lowering transition. Much purer specimen of $RbAgF_3$ and likely much lower temperature, is needed for such detection. The only unambiguous phenomenon observed is that of the thermal expansion; for example, tetragonal lattice constant of $RbAgF_3$ expands by 0.18 % upon warming while the perpendicular lattice vector by 0.33 %; such anisotropy is, obviously, not unexpected, given that ***b*** is axis of propagation of the $AgF_3^-$ infinite chains which ***a*** and ***c*** are equivalent.

### 4.4. Crystal structure and magnetic properties of the Ruddlsden-Popper phase $Rb_3Ag_2F_7$.

One of the studied samples of $RbAgF_3$ showed significant proportion of an impurity phase. This additional phase was not only seen in PXRD measurements, but also during Raman experiments. Since Raman spectrometers equipped with microscope collect data from only a small micron-sized surface spot, it is sometimes possible to identify different phases present in the samples. We have managed, for example, to observe a distinct feature located at 504–507 $cm^{-1}$ (*cf.* ESI). This feature was above the range of the theoretically predicted stiffest Raman-active phonon (477 $cm^{-1}$), but its wavenumber might correspond to the Ag–F stretching of slightly shorter Ag–F bonds than those present in $RbAgF_3$. This fact, together with the substantial number of unassigned reflections in the X-ray diffraction pattern, led us to more detailed study of the new phase. Indexation of the unassigned reflections yielded a strongly elongated tetragonal unit cell. Structural database search pointed out at a similar phase of the $K_3Ag_2F_7$ stoichiometry, which some of us first detected two decades ago. This phase was subsequently refined in the *Ccca* space group.[7] Additional research conducted over the next decade shown that the *Ccca* cell is certainly erroneous since it contains Ag(II) sites in an compressed rather than usual elongated octahedral environment.

A very similar conclusion was reached with respect to $K_2AgF_4$ double perovskite.[7,8,13] Equipped with this knowledge, and aided by DFT+U calculations, we have been able to provide a *Cmca* starting model for the novel $Rb_3Ag_2F_7$ phase. This model was then used in Rietveld refinement of that pattern (full fit and crystallographic information file are available in ESI) and resulted in the structure which closely matched a theoretical one (*cf.* ESI for all theoretically derived .cif files).

Orthorhombic $Rb_3Ag_2F_7$ is a typical Ruddlesden-Popper phase (Figure 12), *i.e.* an intergrowth of the $RbAgF_3$ and $Rb_2AgF_4$ end members of the series. Its formation, while trying to prepare $RbAgF_3$, may be explained by partial reduction of Ag(II)$F_2$ to Ag(I)F upon reaction with walls of the reactor. In consequence, some RbF is left unreacted but its amount does not suffice for the formation of $Rb_2AgF_4$. Instead, a 50:50 compound is formed. This structure can be described as infinite [$AgF_2$] double (sandwich) layers interleaved with rubidium fluoride layers. Each pair of layers is rotated by 45° in respect to the neighbouring ones. The [$AgF_6$] octahedra are distorted in a way already known from $RbAgF_3$ and $Rb_2AgF_4$. Apical fluorides have varying atomic distances to Ag(II) of 2.19(3) and 2.104(4) Å (the shortest distance is that in between sandwich layers). There are two different equatorial Ag–F distances of 2.20(3) and 2.35(3) Å. Apical fluorides form a straight link between Ag sites, but the layers are buckled with the intrasheet Ag–F–Ag angle of 163.4(14)° angle (theoretical value 177°).

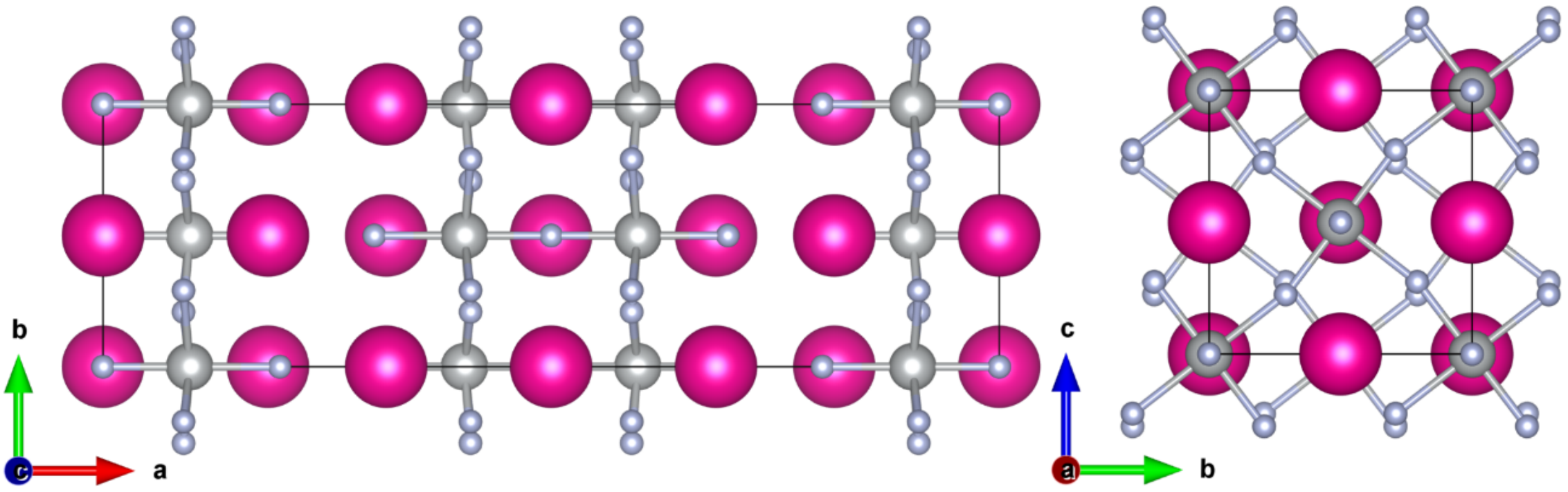

Figure 12. Crystallographic unit cell od novel $Rb_3Ag_2F_7$ phase (pink – rubidium, grey – silver, blue – fluorine).

Table 3. Crystallographic data for novel $Rb_3Ag_2F_7$.

| Formula | $Rb_3Ag_2F_7$ |
|---|---|
| Colour | brown |
| Space group | *Cmca* (No. 64) |
| Z | 4 |
| V /Å$^3$ | 879.524(18) |
| a /Å | 21.7577(5) |
| b /Å | 6.35844(9) |
| c /Å | 6.35747(9) |
| α /° | 90 |
| β /° | 90 |
| γ /° | 90 |
| Positions | Rb1 0.68404(15) 0.00000 0.00000<br>Rb2 0.50000 0.00000 0.00000 0.00776<br>Ag1 -0.09670(16) 0.00000 0.00000<br>F1 0.8028(13) 0.00000 0.00000<br>F2 0.00000 0.00000 0.00000<br>F3 0.5899(7) 0.291(4) 0.274(5) |
| Reliability factors | Not corrected for background (all data)<br>$R_P = 0.84$ %<br>$R_{wp} = 1.78$ %<br><br>Corrected for background (all data)<br>$cR_P = 27.86$ %<br>$cR_{wp} = 15.16$ %<br><br>R Bragg factor (this phase)<br>$R_{Bragg} = 15.81$ % |
| Atomic distances /Å | Ag–F (apical)<br>2.19(3)<br>2.104(4)<br><br>Ag–F (equatorial)<br>2.20(3)<br>2.35(3) |

The shortest Ag–F distance of 2.104(4) Å is slightly shorter than the shortest distance of 2.10935(3) Å seen in tetragonal $RbAgF_3$ at a room temperature. This, together with the different reduced mass factor, might explain why the corresponding oscillator for $Rb_3Ag_2F_7$ is stiffer by 5% than that seen for $RbAgF_3$, as explained above.

Consideration of Goodenough–Kanamori–Anderson rules leads us to the notion, that magnetic interaction between two $[AgF_2]$ layers within a sandwich must be very strong and antiferromagnetic. Indeed, the computed magnetic ground state of $Rb_3Ag_2F_7$ turns out to feature an antiferromagnetic spin ordering in $[Ag_2F_7]$ dimers, ferromagnetic within $[AgF_2]$ layers, and also weakly ferromagnetic between sandwich layers (*i.e.* in a (AB)(BA) manner). Because of that, we have simplified calculations by assuming two dominant superexchange pathways: Ag–F–Ag within a dimer along crystallographic ***a*** axis (this coupling constant is labelled $J_{0D}$), and an intralayer Ag–F…Ag one (labelled $J_{perp}$). Superexchange constants are expressed by equations:

$$J_{0D} = \frac{E_{A2} - E_{A1}}{8}, \quad J_{perp} = \frac{E_{A1} - E_{FM}}{2}$$

($E_{A1}$ is total energy of the ground state, $E_{A2}$ is total energy ordered antiferromagnetically in dimers, antiferomagnetically in layers, $E_{FM}$ is total energy of ferromagnetic solution; energy values per unit cell).

Calculated values are $J_{0D} = -240.2$ meV, $J_{perp} = 10.9$ meV. The $J_{0D}$ value is comparable with those for other strongly coupled Ag(II)-based antiferromagnets calculated to date. Related fluoroargentate(II) systems showing low magnetic dimensionality are $Ag_2ZnZr_2F_{14}$ ($J_{0D} = -313$ meV), HPII-$AgF_2$ ($J_{0D} = -250$ meV) phase,[30] $AgFBF_4$ ($J_{1D} = -298$ meV)[5] and $CsAgF_3$ ($J_{1D} = -161$ meV).[5] On the other hand, $J_{perp}$ is ferromagnetic and comparable to a several meV values typical for $MAgF_3$[6] and $M_2AgF_4$.[8] The ratio of $J_{perp}$ and $J_{0D}$ is *ca.* 0.045 and permits us to label this system as quasi-0D because dimers interact with each other quite weakly.

# 5. Conclusions

Theoretical calculations have led us to conclusion, that at the T → 0 K limit, the previously published structure of $RbAgF_3$[10] is not dynamically stable (in a harmonic approximation). We have been able to propose orthorhombic candidate for the low temperature polymorph, that is isostructural to low-temperature form of $KAgF_3$. Main structural difference between both forms reveals itself in Ag–F–Ag chains, which in tetragonal $RbAgF_3$ are completely straight, and kinked in orthorhombic $RbAgF_3$. There is also an associated puckering of the $[AgF_2]$ sheets in the orthorhombic polytype. Total energy of that orthorhombic phase is 11.7 meV lower than that of the known tetragonal counterpart (HSE06, per Z=4 unit cell), which corresponds to 0.3 kJ $mol^{-1}$ or 34 K.

In order to experimentally confirm a possible existence of low temperature $RbAgF_3$ polymorph, low temperature Raman and powder X-ray diffraction study were conducted. Unfortunate, both methods were inconclusive. Colling to liquid helium temperatures and preparing of a purer specimen is needed to confirm or refute the postulated phase transition. Orthorhombic $RbAgF_3$ should be searched for at LHe temperatures.

Comparative theoretical analysis of lattice dynamics of $MAgF_3$ (M = K, Rb, Cs) compounds was performed. Phonon dispersion curves, phonon density of states and full list of vibrational modes were calculated for orthorhombic $RbAgF_3$ and tetragonal $CsAgF_3$ as a development of our recent study for the K analogue.[14]

Analysis of phonon density of stated graph yielded conclusion, that having larger cation in interstitial positions in a distorted perovskite lattice causes significant localization of certain phonons. We have also compared corresponding theoretically predicted modes to understand trends observed in that series of compounds.

Although at 0 K the orthorhombic $RbAgF_3$ is more stable, in order to analyse room temperature spectral data it was necessary to calculate normal modes for tetragonal structure. Using that, and predicted phonons for $CsAgF_3$ we have assigned specific vibrations to most bands present at IR and Raman spectra. We have reached very good accuracy of the correlation between experimental and theoretical data, thus supporting our assignment.

One of analysed samples of $RbAgF_3$ contained significant portion of contamination phase. Upon indexation, DFT+U modelling, and subsequent Rietveld refinement we have proposed a novel $Rb_3Ag_2F_7$ phase. This new compound constitutes the second known Ruddlesden-Popper phase in fluoroargentate(II) family. $Rb_3Ag_2F_7$ crystallizes in *Cmca* (No. 64) space group (21.7577(5) 6.35844(9) 6.35747(9) 90 90 90). Its structure best described as sandwich layers of [$AgF_2$] stacked at 45°, interleaved with rubidium cations. According to the DFT calculations, this compound features one of the shortest Ag–F bonds known, and it hosts a very strong antiferromagnetic superexchange ($J_{0D}$ = –240.2 meV) within the sandwich layers.

## 6. Acknowledgements

Powder X-ray diffraction measurements were carried out on The Jan Czochralski Laboratory For Advance Crystal Engineering equipment.

Polish authors are grateful to NCN for support (Maestro, 2017/26/A/ST5/00570). Slovenian authors acknowledge the Slovenian Research Agency for financial support within research program P1-0045 Inorganic Chemistry and Technology. Calculations were performed using ICM UW supercomputing resources (okeanos, projects GA76-19 and GA83-34).

## 7. Notes and references

Further details of the crystal structures reported here may be obtained from Fachinformationszentrum Karlsruhe (76344 Eggenstein-Leopoldshafen, Germany; fax (+49)7247-808-666; e-mail: crysdata@fizkarlsruhe.de) on quoting their CSD numbers: 2122449 for $RbAgF_3$ and 2122450 for $Rb_3Ag_2F_7$.

‡ Ionic radius of $MAgF_3$ (M = K, Rb, Cs) compounds increases in the series starting from 1.64 Å for potassium, through 1.72 Å for rubidium, to 1.88 Å for caesium (all for coordination number XII).

1 J. Gawraczyński, D. Kurzydłowski, W. Gadomski, Z. Mazej, G. Ruani, I. Bergenti, T. Jaroń, A. Ozarowski, S. Hill, P. J. Leszczyński, K. Tokár, M. Derzsi, P. Barone, K. Wohlfeld, J. Lorenzana and W. Grochala, *Proc. Natl. Acad. Sci.*, 2019, **116**, 1495–1500.

2 W. Grochala, R. G. Egdell, P. P. Edwards, Z. Mazej and B. Žemva, *ChemPhysChem*, 2003, **4**, 997–1001.

3 N. Bachar, K. Koteras, J. Gawraczynski, W. Trzcinski, J. Paszula, R. Piombo, P. Barone, Z. Mazej, G. Ghiringhelli, A. Nag, K.-J. Zhou, J. Lorenzana, D. van der Marel and W. Grochala, .

4 S. E. McLain, M. R. Dolgos, D. A. Tennant, J. F. C. Turner, T. Barnes, T. Proffen, B. C. Sales and R. I. Bewley, *Nat. Mater.*, 2006, **5**, 561–565.

5 D. Kurzydłowski and W. Grochala, *Angew. Chemie - Int. Ed.*, 2017, **56**, 10114–10117.

6 D. Kurzydłowski and W. Grochala, *Phys. Rev. B*, 2017, **96**, 155140.

7 Z. Mazej, E. Goreshnik, Z. Jagliić, B. Gaweł, W. Łasocha, D. Grzybowska, T. Jaroń, D. Kurzydłowski, P. Malinowski, W. Koźminski, J. Szydłowska, P. Leszczyński and W. Grochala, *CrystEngComm*, 2009, **11**, 1702–1710.

8 D. Kurzydłowski, M. Derzsi, Z. Mazej and W. Grochala, *Dalt. Trans.*, 2016, **45**, 16255–16261.

9 D. Kurzydłowski, Z. Mazej and W. Grochala, *Dalt. Trans.*, 2013, **42**, 2167–2173.

10 R. H. Odenthal and R. Hoppe, *Monatshefte für Chemie*, 1971, **102**, 1340–1350.

11 D. Kurzydłowski, Z. Mazej, Z. Jagličić, Y. Filinchuk and W. Grochala, *Chem. Commun.*, 2013, **49**, 6262–6264.

12 A. I. Popov and Y. M. Kiselev, *Zhurnal Neorg. Khimii*, 1987, **32**, 2276.

13 W. Grochala, 2001.

14 K. Koteras, J. Gawraczyński, M. Derzsi, Z. Mazej and W. Grochala, *Chemistry (Easton).*, 2021, **3**, 94–103.

15 D. Kurzydłowski, T. Jaroń, A. Ozarowski, S. Hill, Z. Jagličić, Y. Filinchuk, Z. Mazej and W. Grochala, *Inorg. Chem.*, 2016, **55**, 11479–11489.

16 J. Gawraczyński, PhD Thesis, University of Warsaw, 2019.

17 H. M. Rietveld, *J. Appl. Crystallogr.*, 1969, **2**, 65–71.

18 J. Rodríguez-Carvajal, .

19 A. V. Krukau, O. A. Vydrov, A. F. Izmaylov and G. E. Scuseria, *J. Chem. Phys.*, 2006, **125**, 224106.

20 G. I. Csonka, J. P. Perdew, A. Ruzsinszky, P. H. T. Philipsen, S. Lebègue, J. Paier, O. A. Vydrov and J. G. Ángyán, *Phys. Rev. B - Condens. Matter Mater. Phys.*, 2009, **79**, 155107.

21 P. E. Blöchl, *Phys. Rev. B*, 1994, **50**, 17953–17979.

22 G. Kresse and D. Joubert, *Phys. Rev. B - Condens. Matter Mater. Phys.*, 1999, **59**, 1758–1775.

23 G. Kresse and J. Hafner, *Phys. Rev. B*, 1993, **47**, 558–561.

24 G. Kresse and J. Furthmüller, *Phys. Rev. B - Condens. Matter Mater. Phys.*, 1996, **54**, 11169–11186.

25 G. Kresse and J. Furthmüller, *Comput. Mater. Sci.*, 1996, **6**, 15–50.

26 X. Zhang, G. Zhang, T. Jia, Y. Guo, Z. Zeng and H. Q. Lin, *Phys. Lett. Sect. A Gen. At. Solid State Phys.*, 2011, **375**, 2456–2459.

27 A. Togo and I. Tanaka, *Scr. Mater.*, 2015, **108**, 1–5.

28 S. Curtarolo, W. Setyawan, G. L. W. Hart, M. Jahnatek, R. V. Chepulskii, R. H. Taylor, S. Wang, J. Xue, K. Yang, O. Levy, M. J. Mehl, H. T. Stokes, D. O. Demchenko and D. Morgan, *Comput. Mater. Sci.*, 2012, **58**, 218–226.

29 R. D. Shannon, *Acta Crystallogr. Sect. A*, 1976, **32**, 751–767.

30 J. Lorenzana, A. Grzelak, V. Struzhkin, D. Kurzydłowski, M. Derzsi, W. Grochala and P. Barone, *Chem. Commun.*, 2018, **54**, 10252–10255.

Electronic supplementary information

# Crystal structure, lattice dynamics and superexchange in $MAgF_3$ 1D antiferromagnets (M = K, Rb, Cs) and $Rb_3Ag_2F_7$ Ruddlesden-Popper phase

Kacper Koteras,[1]* Jakub Gawraczyński,[1] Gašper Tavčar,[2] Zoran Mazej,[2] Wojciech Grochala[1]

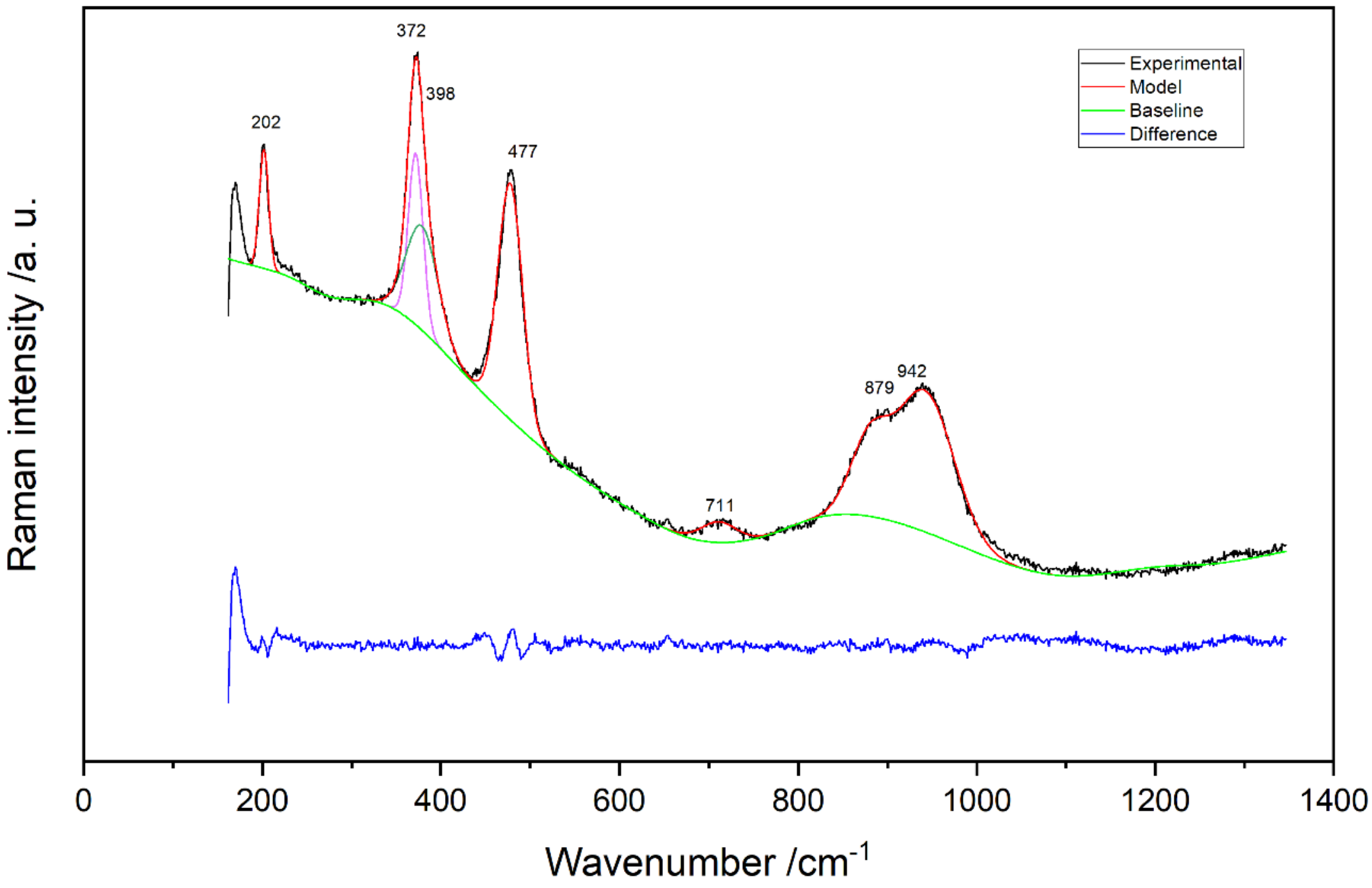

Figure S1. Deconvoluted Raman spectrum of $CsAgF_3$.

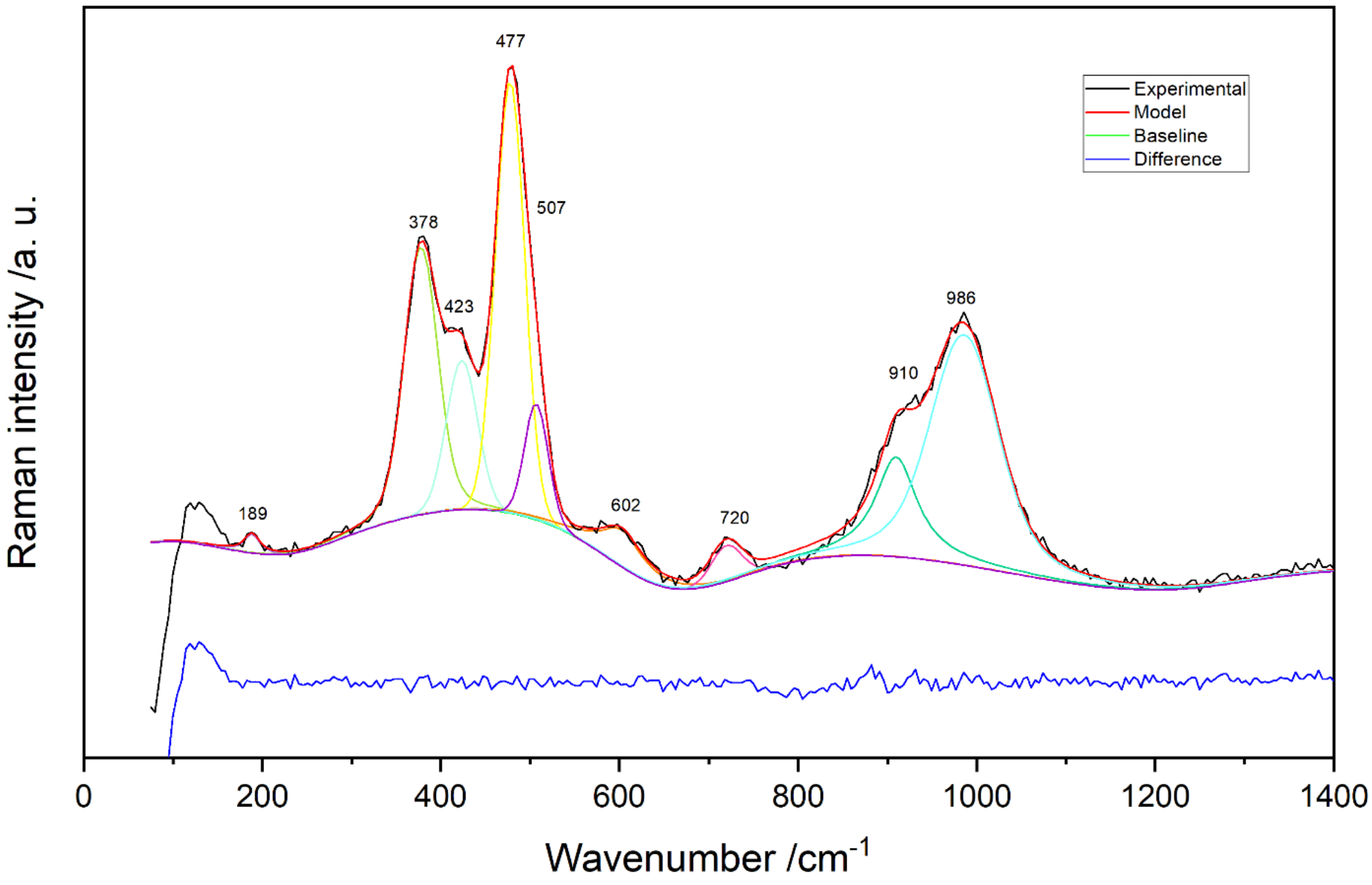

Figure S2. Deconvoluted Raman spectrum of $RbAgF_3$.

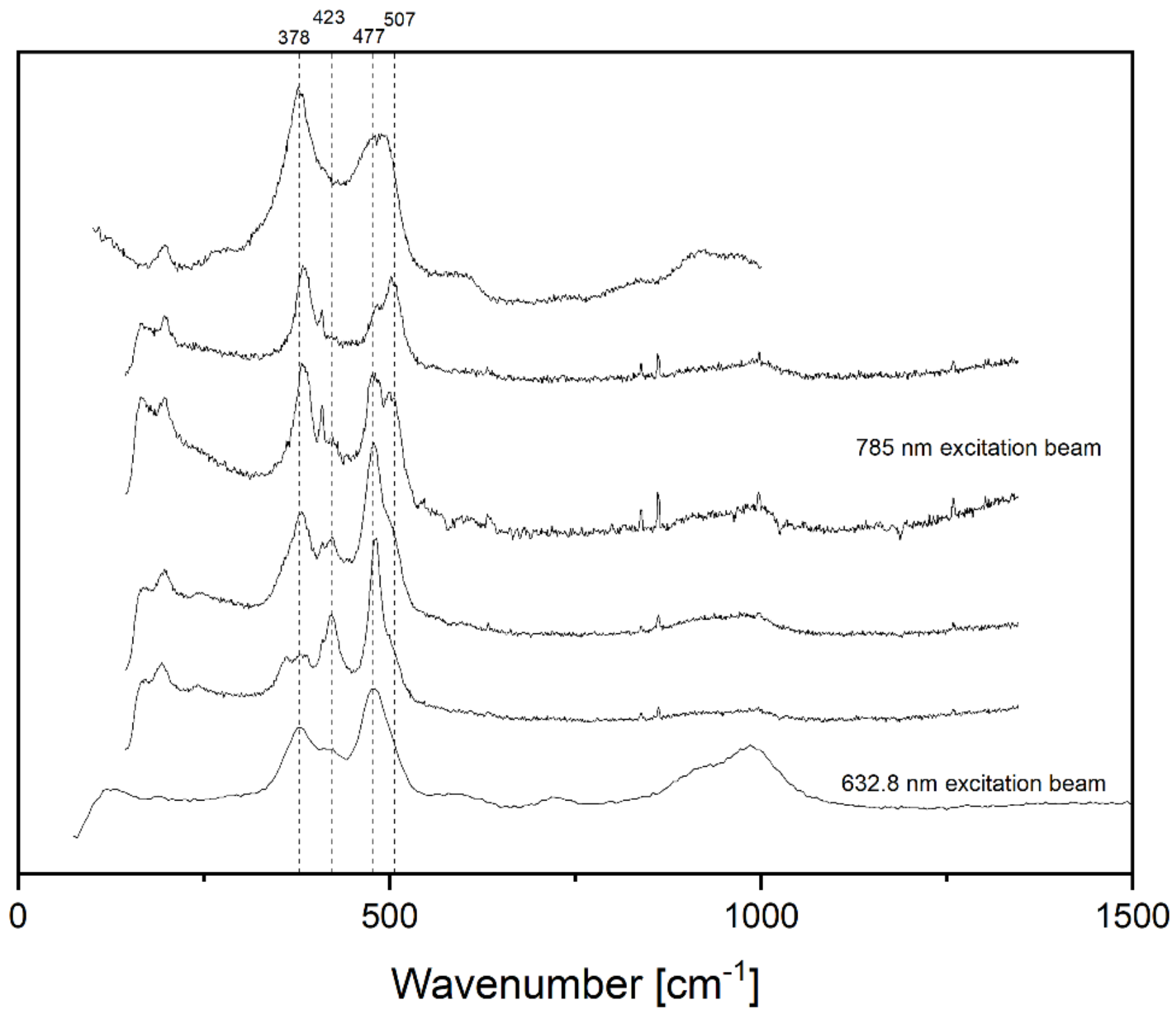

Figure S3. Different Raman spectra collected for of $RbAgF_3$.

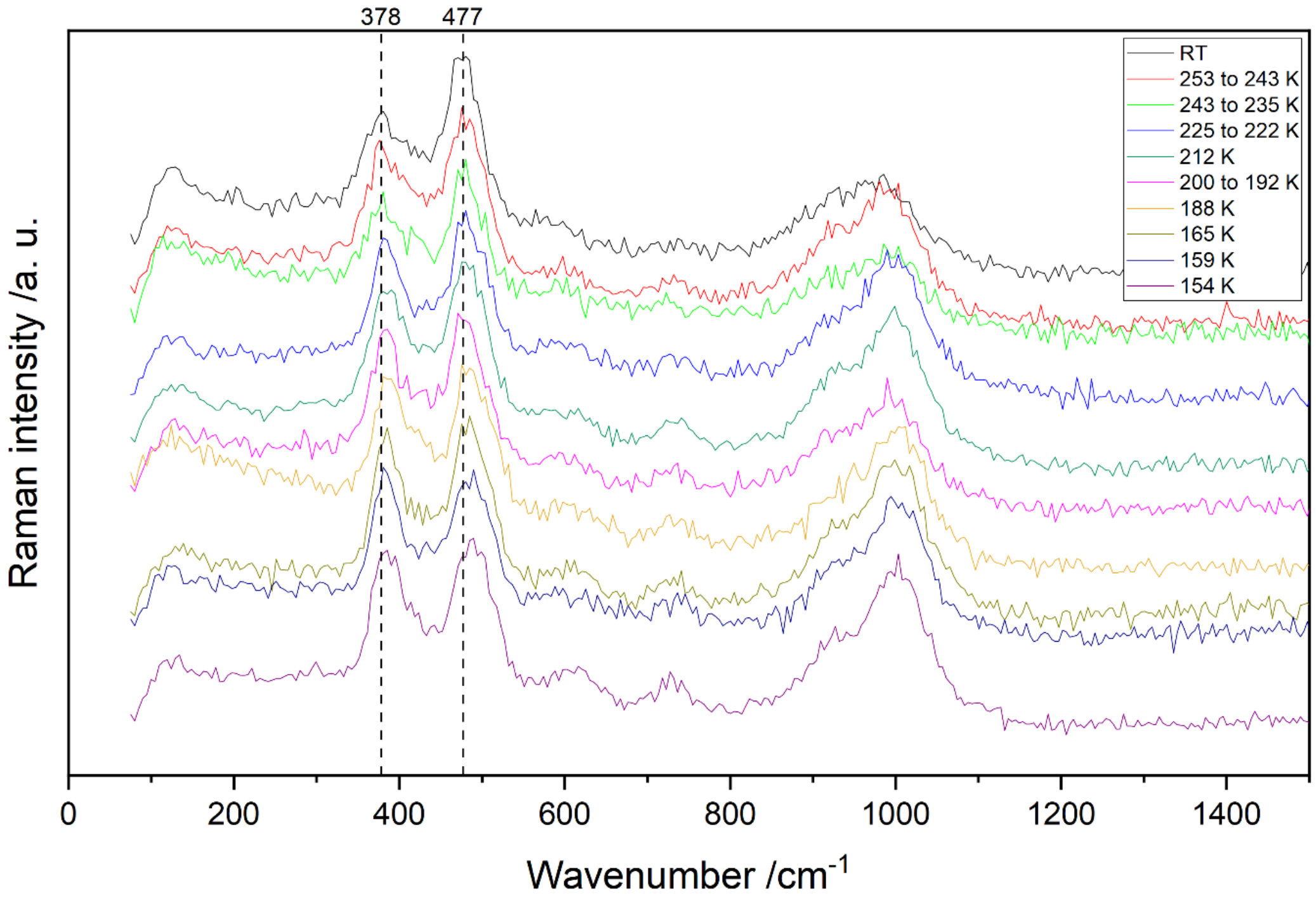

Figure S4. Full set of Raman spectra at low temperature for RbAgF3.

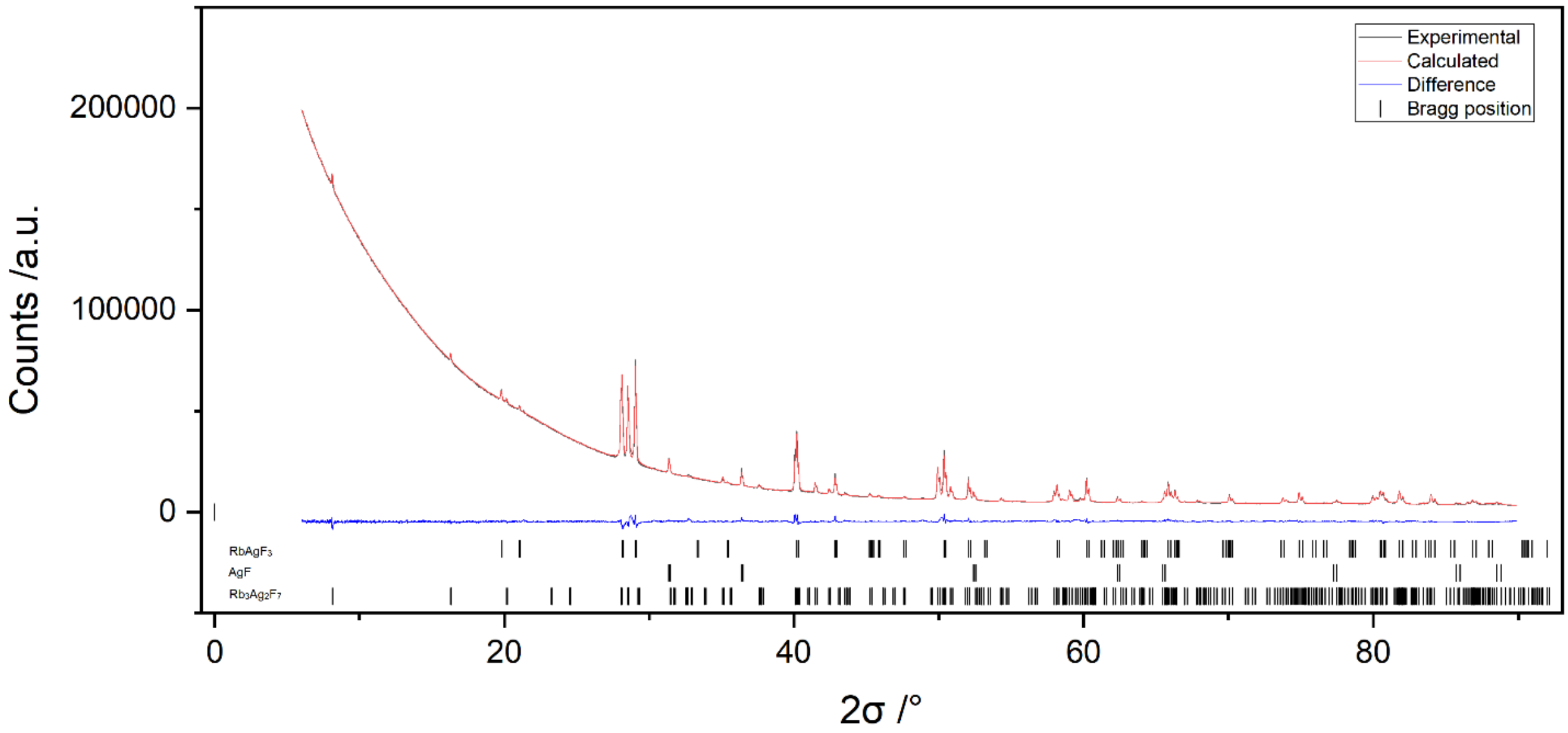

Figure S5. X-ray diffraction pattern of $RbAgF_3$ sample containing novel $Rb_3Ag_2F_7$ phase and the results of the Rietveld refinement. Bottom part shows simulated Bragg positions for each present phase (from top to bottom: $RbAgF_3$, AgF and $Rb_3Ag_2F_7$).

Table S1. Crystallographic data for $RbAgF_3$.

| Formula | $RbAgF_3$ |
|---|---|
| Colour | brown |
| Space group | *I*4/*mcm* (No. 140) |
| Z | 4 |
| V /Å$^3$ | 338.991(6) |
| a /Å | 6.33855(6) |
| b /Å | 6.33855(6) |
| c /Å | 8.43740(11) |
| α /° | 90 |
| β /° | 90 |
| γ /° | 90 |
| Positions | Rb1 0.00000 0.00000 0.25000<br>Ag1 0.00000 0.50000 0.00000<br>F1 0.00000 0.50000 0.25000<br>F2 0.240(4) 0.740(4) 0.00000 |
| Reliability factors | Not corrected for background (all data)<br>$R_P$ = 0.84 %<br>$R_{wp}$ = 1.78 %<br><br>Corrected for background (all data)<br>$cR_P$ = 27.86 %<br>$cR_{wp}$ = 15.16 %<br><br>R Bragg factor (this phase)<br>$R_{Bragg}$ = 9.53 % |
| Atomic distances /Å | Ag–F (apical)<br>2.10935(3)<br><br>Ag–F (equatorial)<br>2.33(3)<br>2.15(3) |

Table S2. Theoretically calulcated modes.

| $RbAgF_3$ *Pnma* (HSE06) | | $RbAgF_3$ *I*4/*mcm* (HSE06) | | $CsAgF_3$ *I*4/*mcm* (HSE06) | |
|---|---|---|---|---|---|
| $B_{2u}$ | 471 | $A_{2u}$ | 484 | $E_u$ | 450 |
| $A_u$ | 470 | $B_{2g}$ | 472 | $B_{2g}$ | 449 |
| $B_{3g}$ | 467 | $E_u$ | 467 | $A_{2u}$ | 449 |
| $B_{1u}$ | 464 | $A_{1g}$ | 381 | $A_{1g}$ | 370 |
| $B_{3u}$ | 464 | $B_{2g}$ | 312 | $B_{2g}$ | 302 |
| $A_u$ | 462 | $B_{1u}$ | 233 | $B_{1u}$ | 227 |
| $B_{2u}$ | 461 | $E_g$ | 184 | $A_{2u}$ | 193 |
| $B_{2g}$ | 449 | $A_{2u}$ | 183 | $E_u$ | 186 |
| $B_{1g}$ | 414 | $E_u$ | 181 | $E_g$ | 186 |
| $A_g$ | 380 | $E_u$ | 159 | $E_u$ | 162 |
| $B_{1g}$ | 365 | $A_{2g}$ | 144 | $B_{1u}$ | 155 |
| $B_{3g}$ | 303 | $B_{1u}$ | 139 | $A_{2g}$ | 150 |
| $B_{3u}$ | 271 | $E_u$ | 124 | $E_u$ | 118 |
| $B_{1u}$ | 240 | $A_{2g}$ | 69 | $B_{1g}$ | 95 |
| $B_{3u}$ | 205 | $A_{2u}$ | 67 | $E_g$ | 88 |
| $A_g$ | 201 | $E_u$ | 66 | $A_{2u}$ | 72 |
| $B_{2g}$ | 200 | $E_g$ | 65 | $E_u$ | 72 |
| $B_{1u}$ | 199 | $E_g$ | 26 | $A_{2g}$ | 65 |
| $B_{2u}$ | 186 | $A_{2u}$ | -1 | $E_g$ | 56 |
| $B_{2g}$ | 182 | $E_u$ | -2 | $A_{2u}$ | -5 |
| $B_{1u}$ | 175 | $B_{1g}$ | -39 | $E_u$ | -5 |
| $B_{3u}$ | 174 | | | | |
| $A_u$ | 174 | | | | |
| $A_u$ | 164 | | | | |
| $B_{2u}$ | 162 | | | | |
| $B_{3u}$ | 161 | | | | |
| $B_{1u}$ | 157 | | | | |
| $B_{3g}$ | 153 | | | | |
| $B_{2u}$ | 146 | | | | |
| $A_u$ | 146 | | | | |
| $B_{1g}$ | 139 | | | | |
| $B_{1u}$ | 139 | | | | |
| $B_{3g}$ | 134 | | | | |
| $B_{3u}$ | 134 | | | | |
| $B_{3u}$ | 129 | | | | |
| $B_{2g}$ | 125 | | | | |
| $B_{1u}$ | 118 | | | | |
| $A_g$ | 98 | | | | |
| $A_g$ | 87 | | | | |
| $B_{1g}$ | 86 | | | | |
| $B_{1u}$ | 82 | | | | |
| $A_g$ | 78 | | | | |
| $B_{3g}$ | 77 | | | | |
| $A_u$ | 76 | | | | |
| $B_{3u}$ | 76 | | | | |
| $B_{2u}$ | 74 | | | | |
| $B_{2g}$ | 73 | | | | |
| $B_{2u}$ | 71 | | | | |
| $A_g$ | 69 | | | | |
| $B_{3u}$ | 68 | | | | |
| $B_{1u}$ | 65 | | | | |
| $A_u$ | 64 | | | | |
| $B_{2g}$ | 60 | | | | |
| $B_{1g}$ | 50 | | | | |
| $A_u$ | 49 | | | | |
| $B_{2g}$ | 34 | | | | |
| $A_g$ | 33 | | | | |
| $B_{2u}$ | -1 | | | | |
| $B_{1u}$ | -1 | | | | |
| $B_{3u}$ | -2 | | | | |

Cif files of structures used in lattice dynamics calculations

**$RbAgF_3$ *Pnma* (HSE06)**

```
data_findsym-output
_audit_creation_method FINDSYM

_cell_length_a    6.2633870000
_cell_length_b    8.3574660000
_cell_length_c    6.3444310000
_cell_angle_alpha 90.0000000000
_cell_angle_beta  90.0000000000
_cell_angle_gamma 90.0000000000
_cell_volume      332.1058636297

_symmetry_space_group_name_H-M "P 21/n 21/m 21/a"
_symmetry_Int_Tables_number 62
_space_group.reference_setting '062:-P 2ac 2n'
_space_group.transform_Pp_abc a,b,c;0,0,0

loop_
_space_group_symop_id
_space_group_symop_operation_xyz
1 x,y,z
2 x+1/2,-y+1/2,-z+1/2
3 -x,y+1/2,-z
4 -x+1/2,-y,z+1/2
5 -x,-y,-z
6 -x+1/2,y+1/2,z+1/2
7 x,-y+1/2,z
8 x+1/2,y,-z+1/2

loop_
_atom_site_label
_atom_site_type_symbol
_atom_site_symmetry_multiplicity
_atom_site_Wyckoff_label
_atom_site_fract_x
_atom_site_fract_y
_atom_site_fract_z
_atom_site_occupancy
_atom_site_fract_symmform
Ag1 Ag   4 a 0.00000 0.00000  0.00000 1.00000 0,0,0
Rb1 Rb   4 c 0.00381 0.25000  0.49913 1.00000 Dx,0,Dz
F1  F    4 c 0.49600 0.25000  0.45836 1.00000 Dx,0,Dz
F2  F    8 d 0.26645 -0.02118 0.27814 1.00000 Dx,Dy,Dz
```

**$RbAgF_3$ *I*4/*mcm* (HSE06)**

```
data_findsym-output
_audit_creation_method FINDSYM

_cell_length_a    6.2963000000
_cell_length_b    6.2963000000
_cell_length_c    8.4031760000
_cell_angle_alpha 90.0000000000
_cell_angle_beta  90.0000000000
_cell_angle_gamma 90.0000000000
_cell_volume      333.1304144144

_symmetry_space_group_name_H-M "I 4/m 2/c 2/m"
_symmetry_Int_Tables_number 140
_space_group.reference_setting '140:-I 4 2c'
_space_group.transform_Pp_abc a,b,c;0,0,0

loop_
_space_group_symop_id
_space_group_symop_operation_xyz
1 x,y,z
2 x,-y,-z+1/2
```

3 -x,y,-z+1/2
4 -x,-y,z
5 -y,-x,-z+1/2
6 -y,x,z
7 y,-x,z
8 y,x,-z+1/2
9 -x,-y,-z
10 -x,y,z+1/2
11 x,-y,z+1/2
12 x,y,-z
13 y,x,z+1/2
14 y,-x,-z
15 -y,x,-z
16 -y,-x,z+1/2
17 x+1/2,y+1/2,z+1/2
18 x+1/2,-y+1/2,-z
19 -x+1/2,y+1/2,-z
20 -x+1/2,-y+1/2,z+1/2
21 -y+1/2,-x+1/2,-z
22 -y+1/2,x+1/2,z+1/2
23 y+1/2,-x+1/2,z+1/2
24 y+1/2,x+1/2,-z
25 -x+1/2,-y+1/2,-z+1/2
26 -x+1/2,y+1/2,z
27 x+1/2,-y+1/2,z
28 x+1/2,y+1/2,-z+1/2
29 y+1/2,x+1/2,z
30 y+1/2,-x+1/2,-z+1/2
31 -y+1/2,x+1/2,-z+1/2
32 -y+1/2,-x+1/2,z

loop_
_atom_site_label
_atom_site_type_symbol
_atom_site_symmetry_multiplicity
_atom_site_Wyckoff_label
_atom_site_fract_x
_atom_site_fract_y
_atom_site_fract_z
_atom_site_occupancy
_atom_site_fract_symmform
Ag1 Ag 4 d 0.00000 0.50000 0.00000 1.00000 0,0,0
Rb1 Rb 4 a 0.00000 0.00000 0.25000 1.00000 0,0,0
F1 F 4 b 0.00000 0.50000 0.25000 1.00000 0,0,0
F2 F 8 h 0.22870 0.72870 0.00000 1.00000 Dx,Dx,0

### $CsAgF_3$ *I*4/*mcm* (HSE06)

data_findsym-output
_audit_creation_method FINDSYM

_cell_length_a 6.4411370000
_cell_length_b 6.4411370000
_cell_length_c 8.4690630000
_cell_angle_alpha 90.0000000000
_cell_angle_beta 90.0000000000
_cell_angle_gamma 90.0000000000
_cell_volume 351.3665678866

_symmetry_space_group_name_H-M "I 4/m 2/c 2/m"
_symmetry_Int_Tables_number 140
_space_group.reference_setting '140:-I 4 2c'
_space_group.transform_Pp_abc a,b,c;0,0,0

loop_
_space_group_symop_id
_space_group_symop_operation_xyz
1 x,y,z
2 x,-y,-z+1/2
3 -x,y,-z+1/2
4 -x,-y,z
5 -y,-x,-z+1/2

6 -y,x,z
7 y,-x,z
8 y,x,-z+1/2
9 -x,-y,-z
10 -x,y,z+1/2
11 x,-y,z+1/2
12 x,y,-z
13 y,x,z+1/2
14 y,-x,-z
15 -y,x,-z
16 -y,-x,z+1/2
17 x+1/2,y+1/2,z+1/2
18 x+1/2,-y+1/2,-z
19 -x+1/2,y+1/2,-z
20 -x+1/2,-y+1/2,z+1/2
21 -y+1/2,-x+1/2,-z
22 -y+1/2,x+1/2,z+1/2
23 y+1/2,-x+1/2,z+1/2
24 y+1/2,x+1/2,-z
25 -x+1/2,-y+1/2,-z+1/2
26 -x+1/2,y+1/2,z
27 x+1/2,-y+1/2,z
28 x+1/2,y+1/2,-z+1/2
29 y+1/2,x+1/2,z
30 y+1/2,-x+1/2,-z+1/2
31 -y+1/2,x+1/2,-z+1/2
32 -y+1/2,-x+1/2,z

loop_
_atom_site_label
_atom_site_type_symbol
_atom_site_symmetry_multiplicity
_atom_site_Wyckoff_label
_atom_site_fract_x
_atom_site_fract_y
_atom_site_fract_z
_atom_site_occupancy
_atom_site_fract_symmform
Ag1 Ag 4 d 0.00000 0.50000 0.00000 1.00000 0,0,0
Cs1 Cs 4 a 0.00000 0.00000 0.25000 1.00000 0,0,0
F1 F 4 b 0.00000 0.50000 0.25000 1.00000 0,0,0
F2 F 8 h 0.22363 0.72363 0.00000 1.00000 Dx,Dx,0

**$Rb_3Ag_2F_7$ *Cmca* (GGA+U)**

data_findsym-output
_audit_creation_method FINDSYM

_cell_length_a 21.6756500000
_cell_length_b 6.2943600000
_cell_length_c 6.2937100000
_cell_angle_alpha 90.0000000000
_cell_angle_beta 90.0000000000
_cell_angle_gamma 90.0000000000
_cell_volume 858.6781972783

_symmetry_space_group_name_H-M "C 2/m 2/c 21/a"
_symmetry_Int_Tables_number 64
_space_group.reference_setting '064:-C 2ac 2'
_space_group.transform_Pp_abc a,b,c;0,0,0

loop_
_space_group_symop_id
_space_group_symop_operation_xyz
1 x,y,z
2 x,-y,-z
3 -x+1/2,y,-z+1/2
4 -x+1/2,-y,z+1/2
5 -x,-y,-z
6 -x,y,z

7 x+1/2,-y,z+1/2
8 x+1/2,y,-z+1/2
9 x+1/2,y+1/2,z
10 x+1/2,-y+1/2,-z
11 -x,y+1/2,-z+1/2
12 -x,-y+1/2,z+1/2
13 -x+1/2,-y+1/2,-z
14 -x+1/2,y+1/2,z
15 x,-y+1/2,z+1/2
16 x,y+1/2,-z+1/2

loop_
_atom_site_label
_atom_site_type_symbol
_atom_site_symmetry_multiplicity
_atom_site_Wyckoff_label
_atom_site_fract_x
_atom_site_fract_y
_atom_site_fract_z
_atom_site_occupancy
_atom_site_fract_symmform
Ag1 Ag 8 d 0.40284 0.00000 0.00000 1.00000 Dx,0,0
Rb1 Rb 8 d 0.18085 0.00000 0.00000 1.00000 Dx,0,0
Rb2 Rb 4 a 0.00000 0.00000 0.00000 1.00000 0,0,0
F1 F 8 d 0.30625 0.00000 0.00000 1.00000 Dx,0,0
F2 F 4 b 0.50000 0.00000 0.00000 1.00000 0,0,0
F3 F 16 g 0.09527 0.26940 0.23077 1.00000 Dx,Dy,Dz